\documentclass[journal]{IEEEtran}
\ifCLASSINFOpdf
 \usepackage[pdftex]{graphicx}
\else
\fi
\usepackage{cite}
\usepackage{amsmath,amssymb,amsfonts}
\usepackage{txfonts}
\usepackage{bm}
\usepackage{algorithmic}
\usepackage{graphicx}
\usepackage{textcomp}
\usepackage{subfigure}
\usepackage{threeparttable}
\usepackage{float}
\usepackage{epsfig}
\usepackage{epstopdf}
\usepackage{dsfont}
\usepackage{amsmath}
\usepackage{multirow}
\usepackage{color, soul}
\soulregister\cite7 
\soulregister\citep7 
\soulregister\citet7 
\soulregister\ref7 
\soulregister\pageref7 
\newcommand{\tabincell}[2]{\begin{tabular}{@{}#1@{}}#2\end{tabular}}
\def\BibTeX{{\rm B\kern-.05em{\sc i\kern-.025em b}\kern-.08em
    T\kern-.1667em\lower.7ex\hbox{E}\kern-.125emX}}
\usepackage{fancyhdr}

\begin{document}
%
\title{Cluster-based Characterization and Modeling for UAV Air-to-Ground Time-Varying Channels}
%
  \author{Zhuangzhuang~Cui,~\IEEEmembership{Graduate Student Member,~IEEE,}~Ke~Guan,~\IEEEmembership{Senior Member,~IEEE,}\\Claude Oestges,~\IEEEmembership{Fellow,~IEEE},~C\'esar~Briso-Rodr\'iguez,~\IEEEmembership{Member,~IEEE},\\Bo Ai,~\IEEEmembership{Senior Member,~IEEE},~and~Zhangdui~Zhong,~\IEEEmembership{Senior Member,~IEEE} 

  \thanks{This work was supported by the NSFC under Grant (61771036, 61911530260, 61901029, and 61725101), the State Key Laboratory of Rail Traffic Control and Safety (Contract No. RCS2020ZZ005), and the Project of China Shenhua under Grant (GJNY-20-01-1). (\emph{Corresponding author: Ke Guan}).}
  \thanks{Z. Cui, K. Guan, B. Ai, and Z. Zhong are with the State Key Lab of Rail Traffic Control and Safety, Beijing Jiaotong University, Beijing, 100044 China (e-mail: \{cuizhuangzhuang, kguan, bai, zhdzhong,\}@bjtu.edu.cn.)}
    \thanks{C. Oestges is with the ICTEAM Institute, Universit\'e Catholique de Louvain, B-1348 Louvain-la-Neuve, Belgium (e-mail: claude.oestges@uclouvain.be).}
  \thanks{C. Briso-Rodr\'iguez is with the Department of Signal Theory and Communications, Universidad Polit\'ecnica de Madrid, Madrid, 28038 Spain. (e-mail: cesar.briso@upm.es).}}
\markboth{}{Z.~Cui \MakeLowercase{\textit{et al.}}: Cluster-based Characterization and Modeling for UAV Air-to-Ground Time-Varying Channels}

\maketitle

\begin{abstract}
With the deep integration between the unmanned aerial vehicle (UAV) and wireless communication, UAV-based air-to-ground (AG) propagation channels need more detailed descriptions and accurate models. In this paper, we aim to perform cluster-based characterization and modeling for AG channels. To our best knowledge, this is the first study that concentrates on the clustering and tracking of multipath components (MPCs) for time-varying AG channels. Based on measurement data at 6.5~GHz with 500~MHz of bandwidth, we first estimate potential MPCs utilizing the space-alternating generalized expectation-maximization (SAGE) algorithm. Then, we cluster the extracted MPCs considering their static and dynamic characteristics by employing K-Power-Means (KPM) algorithm under multipath component distance (MCD) measure. For characterizing time-variant clusters, we exploit a clustering-based tracking (CBT) method, which efficiently quantifies the survival lengths of clusters. Ultimately, we establish a cluster-based channel model, and validations illustrate the accuracy of the proposed model. This work not only promotes a better understanding of AG propagation channels but also provides a general cluster-based AG channel model with certain extensibility.
\end{abstract}

\section{Introduction}
\IEEEPARstart{W}{ITH} the evolution of the next-generation communication system, non-terrestrial networks (NTNs) have received many interests. Compared to terrestrial communications, such as cellular and vehicular ones, unmanned aerial vehicle (UAV)-empowered aerial communications are becoming more and more popular, thanks to their potentials in broad coverage and easy-to-deployment \cite{r1}. In the emerging communication, the UAV is not only a user of cellular systems but also acts as an aerial base station (ABS) with high flexibility. For example, UAVs utilized as ABSs have been implemented by Huawei \cite{huawei}, Nokia \cite{nokia}, and AT\&T \cite{att1}. In practice, tethered or non-tethered drones equipped with remote radio units (RRUs) of base stations provide temporary or enhanced connectivity to the ground user in emergency or hot-spot scenarios.

For the optimal design and performance of any UAV-based air-to-ground (AG) communication system, the better understanding and accurate modeling of corresponding AG propagation channels are indispensable. However, prior works mainly focus on narrowband channel characteristics and models. For instance, multi-frequency UAV channel measurements were conducted in a campus environment \cite{zc1, cb1}. The AG channels at L-band and C-band were comprehensively measured and modeled for different environments such as over-water, built-up, and hilly areas \cite{dw1,dw2,dw3}. It is acknowledged that the large-scale channel characteristics such as path loss, shadowing, and fading margins are helpful for the preliminary deployment of AG communication systems. However, for a meticulous design such as the symbol length, multipath propagation may lead to adverse effects such as intersymbol interference (ISI) owing to the delay dispersion. Hence, the effort on multipath effects of wideband AG channels is urgently required \cite{qz}. Generally, multipath channels can be interpreted in two different ways: one is the channel transfer function (CTF), and the other is the channel impulse response (CIR). Nonetheless, these two interpretations are equivalent by performing Fourier transformation between delay and frequency domain \cite{tsr1}. In this paper, our objective is to accurately and efficiently model the CIRs of time-varying AG channels by considering the clustering and tracking of multipath components (MPCs).

It is vital to elucidate our motivation for clustering and tracking of MPCs for wideband AG channels. It makes no doubt that it can achieve the highest accuracy to calculate channel parameters with all extracted individual MPCs. However, it is highly complicated and has a limited intuitive description of channel behavior. Comparatively, in the clustering and tracking, we use parameterized formulas to illustrate the propagation behaviors, which can not only reduce the processing complexity but also provide a better understanding of wireless channels \cite{rh1}. In addition to the complexity, the clustering and tracking processes conform with the corresponding physical observations. For instance, some MPCs coming from the same scatterer may form a cluster, and the MPC cluster will appear or disappear with the mobility of the transceiver, which is the essence of the tracking process \cite{tw1}. Besides, the clustering focuses on the delay similarity in a specific distance, whereas the tracking process emphasizes the distance continuity and delay similarity. Thus, the clustering and tracking completely capture the static and dynamic characteristics of MPCs in time-varying channels, which motivates us from a technical perspective.



Retrospectively speaking, the cluster-based channel modeling originates from \cite{hs1} and is further developed as Saleh-Valenzuela (SV) model \cite{sv}, in which the first rays of clusters and sub-rays in a cluster are modeled as Poisson arrival processes with fixed rates. At present, cluster-based channel models have been widely adopted in the standard models such as COST 2100 \cite{cost}, 3GPP Spatial Channel Model \cite{3gpp}, and WINNER II \cite{winner}. Unfortunately, current cluster-based channel models focus primarily on conventional channels such as vehicular channels \cite{my1, ch, hj}, which are quite different from aerial channels. As an example, ground vehicles generally move in a linear trajectory and with fixed height, however, the trajectories and heights of aerial vehicles are highly varying in the three-dimensional (3D) space, which leads to a large discrepancy in the characteristics of MPCs, such as the number, power, existence time, and so on \cite{ccr}.

The popular methods of clustering from machine learning algorithms that concentrate more on the data feature such as degrees of separation and compactness, include Kernel-Power-Density (KPD) \cite{rh1}, K-Power-Means (KPM) \cite{jl1}, and K-Means (KM) \cite{um}. These methods have been widely used in vehicle-to-vehicle (V2V), massive Multi-Input Multi-Output (mMIMO), and outdoor-to-indoor (O2I) scenarios. However, the clustering work regarding with UAV-based AG channel is still in its infancy. Fortunately, the tracking of MPCs for time-varying AG channels was studied in \cite{zyh} by employing an improved multipath component distance (MCD) algorithm.  The birth-and-death process of MPC is defined as \textit{trajectory} of MPC \cite{ch1}, where the trajectory length represents the survival time of specific MPC, which well describes the time-evolved characteristics of MPCs. However, the clustering of MPC is still absent \cite{zyh}. This paper is the first work that focuses on the clustering and tracking of MPCs for UAV-based time-varying AG channels, which aims to fill existing gaps. The main contributions are summarized as follows.
\begin{itemize}
\item We analyze the static and dynamic characteristics of MPCs for time-varying AG channels, including the extraction, clustering, and tracking of MPCs, which provides a better understanding of AG multipath channels.

\item Inter-cluster and intra-cluster characteristics are thoroughly analyzed, including the number of clusters, the power decay function, and the cluster delay modeling, which facilitates assembling the cluster-based channel model.

\item A clustering-based tracking method is firstly proposed based on weighted 3D Euclidean distance, which efficiently describes the time-variant characteristics of clusters under the survival length measure. 

\item We validate the proposed cluster-based model by comparing measured and simulated channel parameters. Moreover, we generate the clustered delay line (CDL) model of the UAV-based AG scenario to fill the gap of the 3GPP channel model.
\end{itemize}

The remainder of this paper is organized as follows. Section II introduces measurement data collection and initially refines the obtained data. Cluster-based channel characterization and modeling are performed in section III. Then, we summarize the procedure of model implementation and validate the proposed channel model in section IV. At last, the discussion and conclusion are drawn in sections V and VI, respectively.

\begin{figure}[!t]
  \centering
 \includegraphics[width=2.6in]{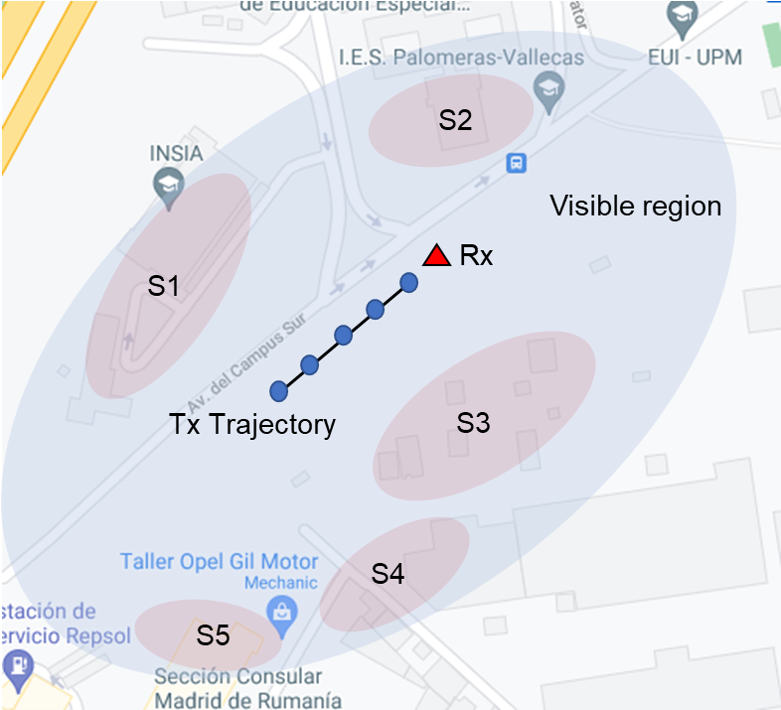}
  \caption{A top view of channel measurement environment with marked possible scatterer groups (S1--S5), Rx, and Tx trajectory in the visible region.}
  \label{maps}
 \end{figure}
\section{Data Collection and Preliminary Processing}
In this section, we will briefly introduce the channel measurement campaign conducted in a campus environment at the Technical University of Madrid, Spain. Then, the preliminary processing of the obtained CIRs will be launched, where the potential MPCs will be extracted by the space-alternating generalized expectation-maximization (SAGE) algorithm.

\subsection{Measurement Campaign}
The AG channel measurements were carried out with commercial ultra-wideband (UWB) modules and DJI UAV, with a central operating frequency $f_c=6.5$ GHz and a bandwidth of 500 MHz \cite{dwm}. The large bandwidth enables to increase the delay resolution so that the most MPCs can be captured. Measurements were conducted in a campus environment that can be regarded as a typical suburban environment according to the density and average height (15~m) of surrounding buildings.  As shown in Fig.~\ref{maps}, we illustrate the location of the receiver (Rx) and the trajectory of the UAV equipped with a transmitter (Tx). The Rx was placed around 0.5 m above the ground level. The UAV flew away from the Rx, with the 3D link distance ranging from 10~m to 50~m. We identify the visible region determined by the maximum detectable delay ($\tau_{\max}=550$ ns). More physically, we also mark the scatterer groups that may be the sources of clusters, where scatterers are mainly composed of buildings (S1, S2, S4, and S5) and large containers (S3). Notably, the antenna patterns of transceivers are approximately omnidirectional. More measurement details can refer to our previous work \cite{zc2}.

\subsection{MPC Estimation}
\begin{figure}[!t]
  \centering
 \subfigure[]{\includegraphics[width=2.8in]{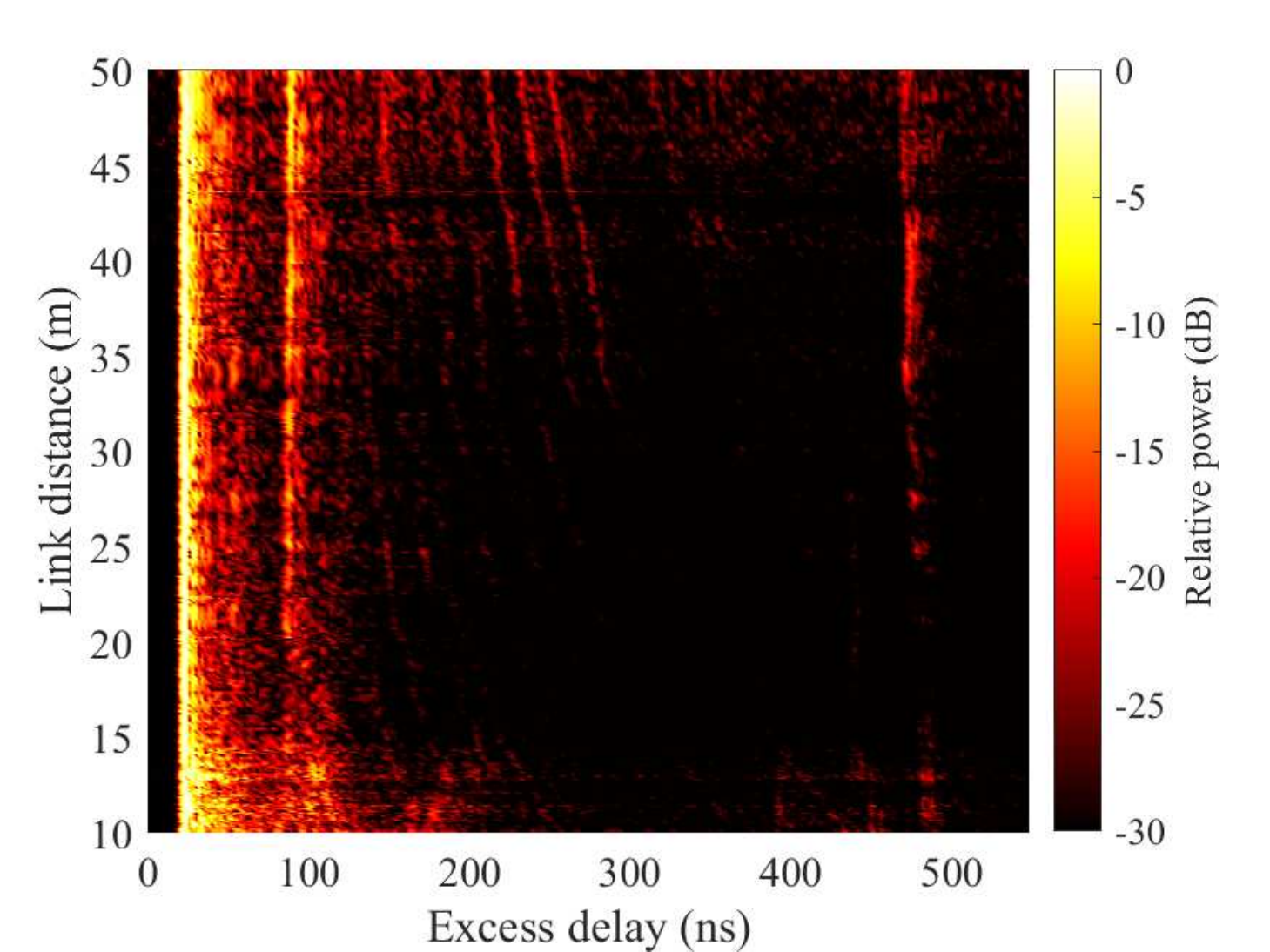}}  \subfigure[]{\includegraphics[width=2.8in]{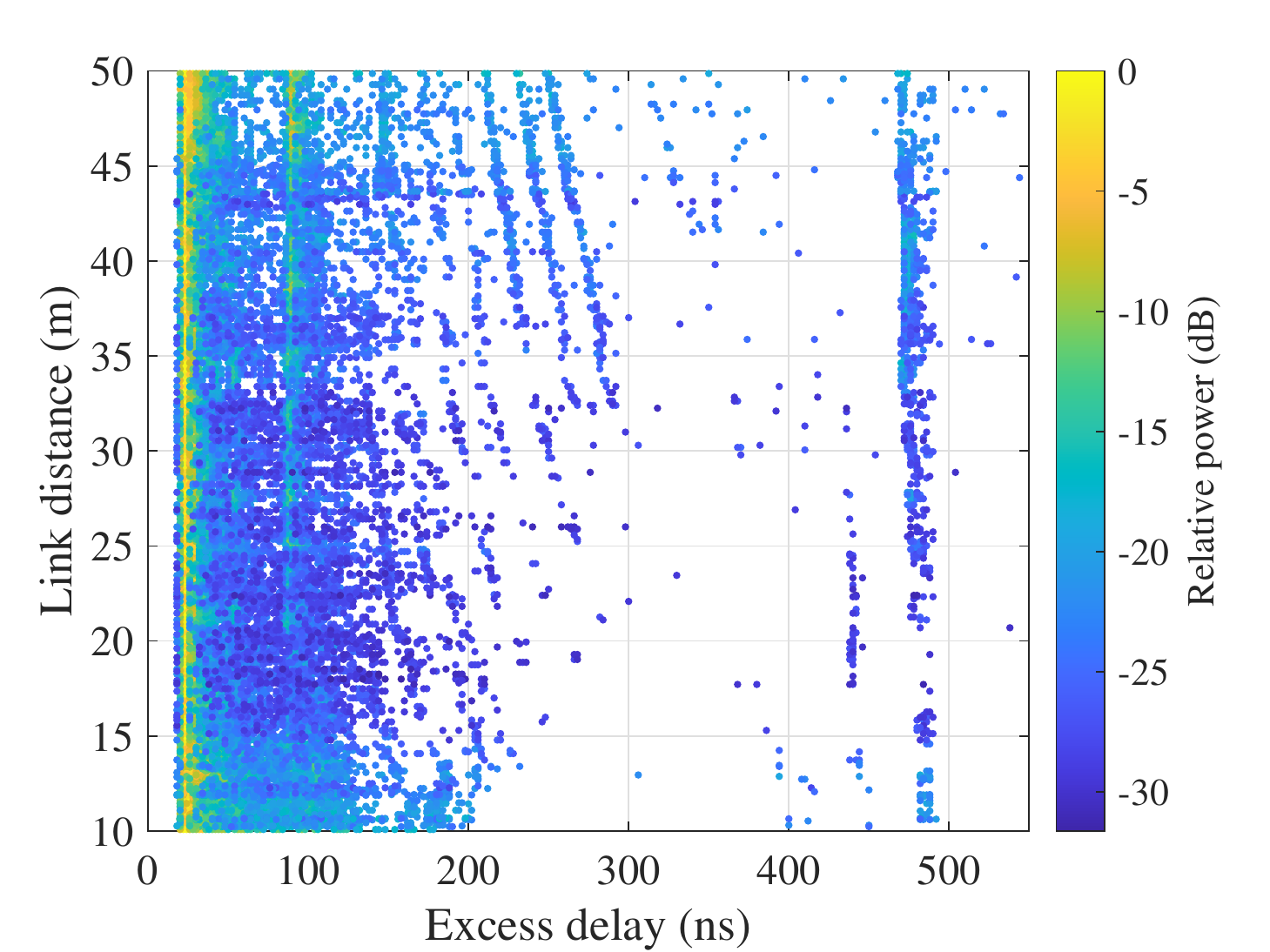}}
  \caption{PDP results: (a) measurement results, (b) extracted MPCs.}
  \label{pdp}
 \end{figure}

Instantaneous power delay profiles (PDPs) can be obtained by the measured CIRs, which are given by $P(t,\tau)=|h(t,\tau)|^2$. Notably, $N$ discrete successive snapshots were stored, and represent the temporal continuity ($t=i\Delta t, i\in\{1, 2, ..., N\}$).  Besides, the transmitted signal is a continuous wave with power of $-17$~dBm, and the Rx sensitivity is $-98$~dBm, which results in a dynamic range of 81 dB. For the preliminary processing, we normalize the power for convenience and keep it ranging from 0 to -30~dB, which indicates that MPCs with power smaller $10^3$ times than the strongest component are preserved. As shown in Fig.~\ref{pdp}(a), we can observe the apparent MPCs in measured PDPs. However, for clustering, it is necessary to extract individual MPCs with power and delay information.

For the estimation of MPC, several high-resolution extraction algorithms are widely used, where we use the SAGE algorithm that is verified high accuracy \cite{xc1, jl1, xw}. More details of the SAGE algorithm can be found in \cite{sage}. Note that since we employed a single antenna for both Tx and Rx, the angular information of MPC is absent from measurements. Hence, the parameter set of estimated MPCs is denoted as $\Omega=[\alpha_l(i), \tau_l(i)]$ representing the complex amplitude and delay of the $l$th path in the $i$th snapshot, respectively. The instantaneous power can be obtained by $P_l =|\alpha_l|^2$. For the initialization of SAGE, we define the number of MPCs as 50, which resides in a reasonable range according to prior works \cite{kg1}. As shown in Fig.~\ref{pdp}(b), we plot the extracted MPCs. Comparing with Fig.~\ref{pdp}(a), it shows that the visually observed potential MPCs are efficiently extracted.


 \section{Cluster-based Channel Characterization and Modeling}
 In this section, we will systematically illustrate the methodology of cluster-based channel modeling and characterization. We will first introduce the clustering method and determine the optimal number of clusters. Then, we will comprehensively investigate the intra-cluster and inter-cluster characteristics. Moreover, the corresponding models in terms of delay, power, number, and birth-and-death of clusters will be proposed to compose the complete cluster-based models of AG channels.
 
\subsection{Time-Varying Channel Description}
In order to describe the time-varying channel, the general representation of cluster-based CIR is given by \cite{cir}
 \begin{equation}
h(t, \tau) = \sum_{k=1}^{K} \sum_{l=1}^{L}\alpha_{k, l}e^{-j\phi_{k,l}}\delta(t-\tau_{k}-\tau_{k,l}),
\end{equation}
where $K$ and $L$ are the numbers of clusters and rays in a cluster, respectively. In particular, $\tau_k$ is the delay of the $k$-th cluster, and $\tau_{k,l}$ is the delay of the $l$-th path in the $k$-th cluster. Finally, $\delta(\cdot)$ is the Dirac delta function, and $\phi_{k,l}$ is the phase of MPC that is assumed to be described by statistically independent random variables uniformly distributed over $[0,2\pi)$.  We herein focus on each individual MPCs and aim to cluster and track them in a proper way. Therefore, the main focus resides on modeling the clustered MPC power and delay, i.e.,  $\{\alpha_{k, l},\tau_{k,l}\}$.



 \subsection{K-Power-Means Clustering}
The KPM \cite{kpm} is an evolved algorithm of the KM method \cite{km}, incorporating the MPC power as the weight. To employ the KPM algorithm, the distance between MPCs, namely, MCD, is firstly calculated in the delay domain, given by
\begin{equation}
	\text{MCD}_{\tau, i, j} = \zeta\cdot \frac{|\tau_i-\tau_j|}{\Delta \tau_{\max}}\cdot\frac{\tau_{\text{std}}}{\Delta \tau_{\max}},
\end{equation}
where $\Delta \tau_{\max}$ and $\tau_{\text{std}}$ are the maximum difference and the standard deviation of the MPC delays, respectively. $\zeta$ represents the delay scaling factor, which is chosen as 1. Here we denote $\textbf{x}=\{\tau_1,\tau_2, ..., \tau_L\}$. The concrete steps of KPM are as follows.

\begin{itemize}
\item [1)] Randomly choose $K$ initial centroid positions, denoted as $\textbf{c}_1^{(0)}$, ..., $\textbf{c}_K^{(0)}$.
\item [2)] Assign MPCs to cluster centroids and store indices,
\begin{equation}
\mathcal{I}_l^e = \arg \min\{P_l \cdot \text{MCD}_{\tau, x_l, \textbf{c}_k^{(e-1)}}\},
\end{equation}
where ($e$) represents the $e$-th iteration.
\item [3)] Update the cluster centroids and denote
\begin{equation}
\textbf{c}_k^{(e+1)}= \frac{\sum_{x\in\textbf{x}} \mathds{1} \{\mathcal{I}_l^e =e\} x\cdot P_l }{\sum_{x\in\textbf{x}} \mathds{1} \{\mathcal{I}_l^e =e\} P_l}.
\end{equation}
\item [4)] Return clusters if $\textbf{c}_k^{(e+1)}=\textbf{c}_k^{(e)}$. Otherwise, repeat steps 2 and 3 until the convergence is achieved.
\end{itemize}

\begin{figure}[!t]
  \centering
 \includegraphics[width=2.8in]{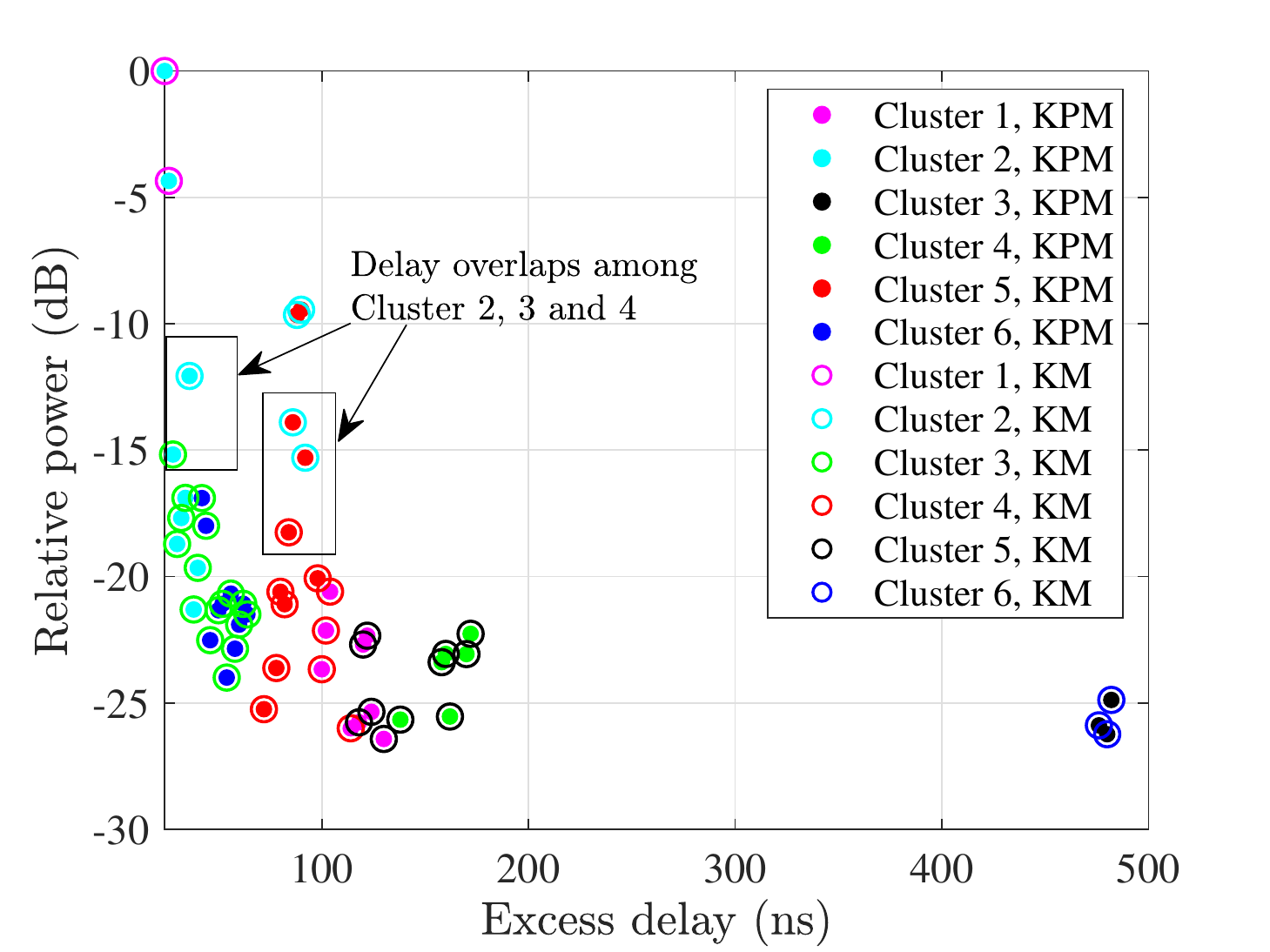}
  \caption{A comparison between KPM and KM clustering results.} 
  \label{kmkpm}
 \end{figure}

As shown in Fig.~\ref{kmkpm}, we compare the clustering results between MCD-based KPM and KM that employs the 2D Euclidean distance defined as $D_{ij}=\sqrt{(\tau_i-\tau_j)^2+(P_i-P_j)^2}$. Results show that employing the KPM algorithm can obtain clusters with successive delays, which means no overlap between clusters in the delay domain. However, several overlaps that we marked in Fig.~\ref{kmkpm} exist under the KM clustering. Thus, for better descriptions of cluster delay and associated power decay function, we utilize the KPM algorithm in the sequel.

\subsection{Number of Cluster}
Note that the number of clusters requires being predefined in the KPM method. Moreover, we need to determine the optimal number for a better clustering performance. Generally, several measures can be used to assess the clustering performance, and thus to find the optimal number, such as Silhouette \cite{si} and Davies-Bouldin (DB) indices \cite{db}. Hereupon, we show the calculation of the DB index as an example, where the compactness $S_k$ is calculated by
\begin{equation}
S_k= \frac{1}{L_k}\sum_{l=1}^{L_k} \text{MCD}(\textbf{x}_l, \textbf{c}_k).
\end{equation}
The separation between two centroids $i$ and $j$ is given by $d_{ij}= \text{MCD}(\textbf{c}_i, \textbf{c}_j)$. Consequently, the DB index is given by
\begin{equation}
\text{DB} (K)= \frac{1}{K}\sum_{i-1}^K R_i,
\end{equation}
with
\begin{equation}
R_i= \max_{j=1...K, j\neq i} \frac{S_i+S_j}{d_{ij}}.
\end{equation}
Thus, the optimal number of clusters can be determined by
\begin{equation}
K_{\rm opt}=\arg\max_{K} \{\text{DB} (K)\}.
\end{equation}

\begin{figure}[!t]
  \centering
\subfigure[]{\includegraphics[width=1.6in]{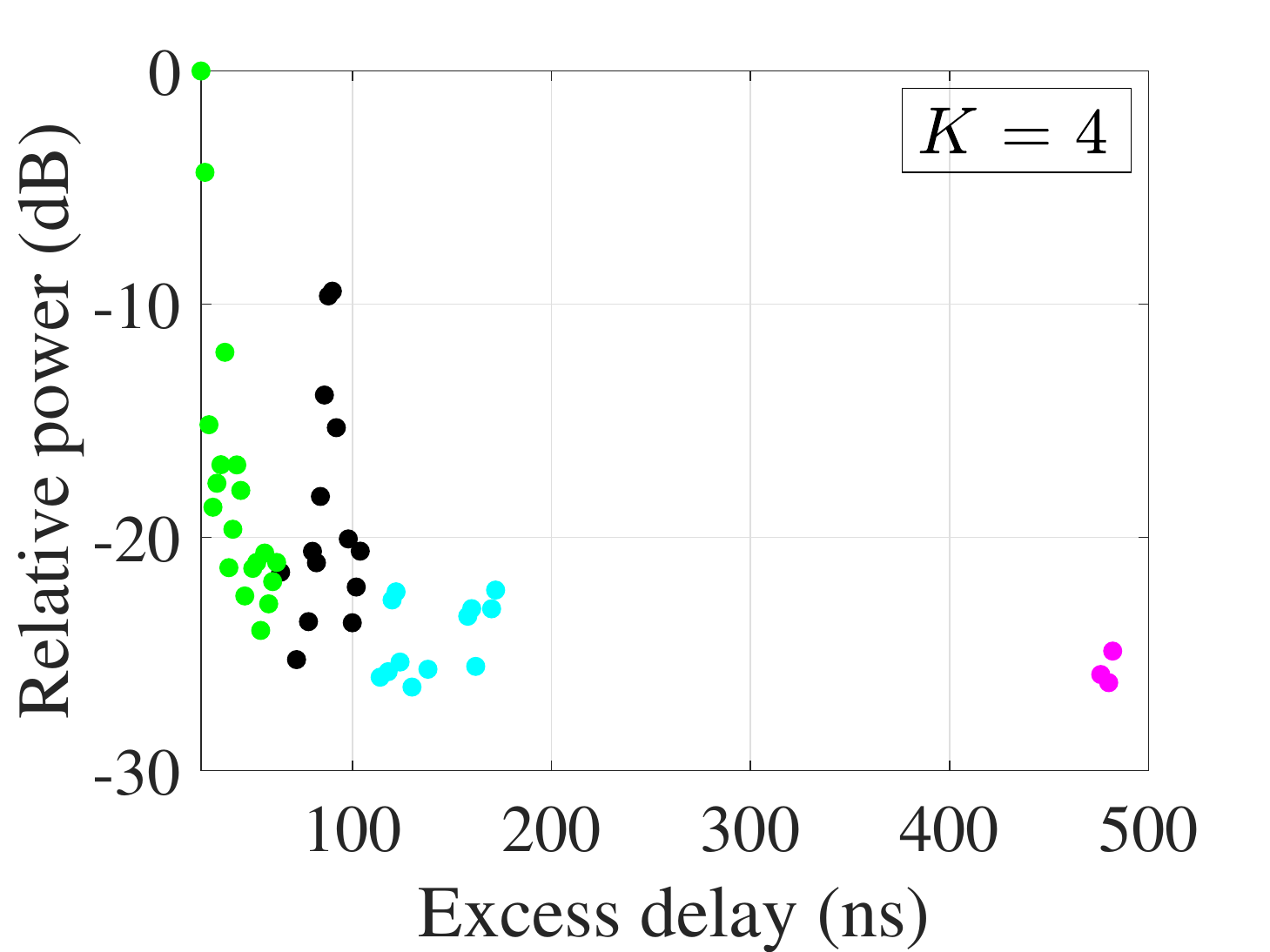}}
  \subfigure[]{\includegraphics[width=1.6in]{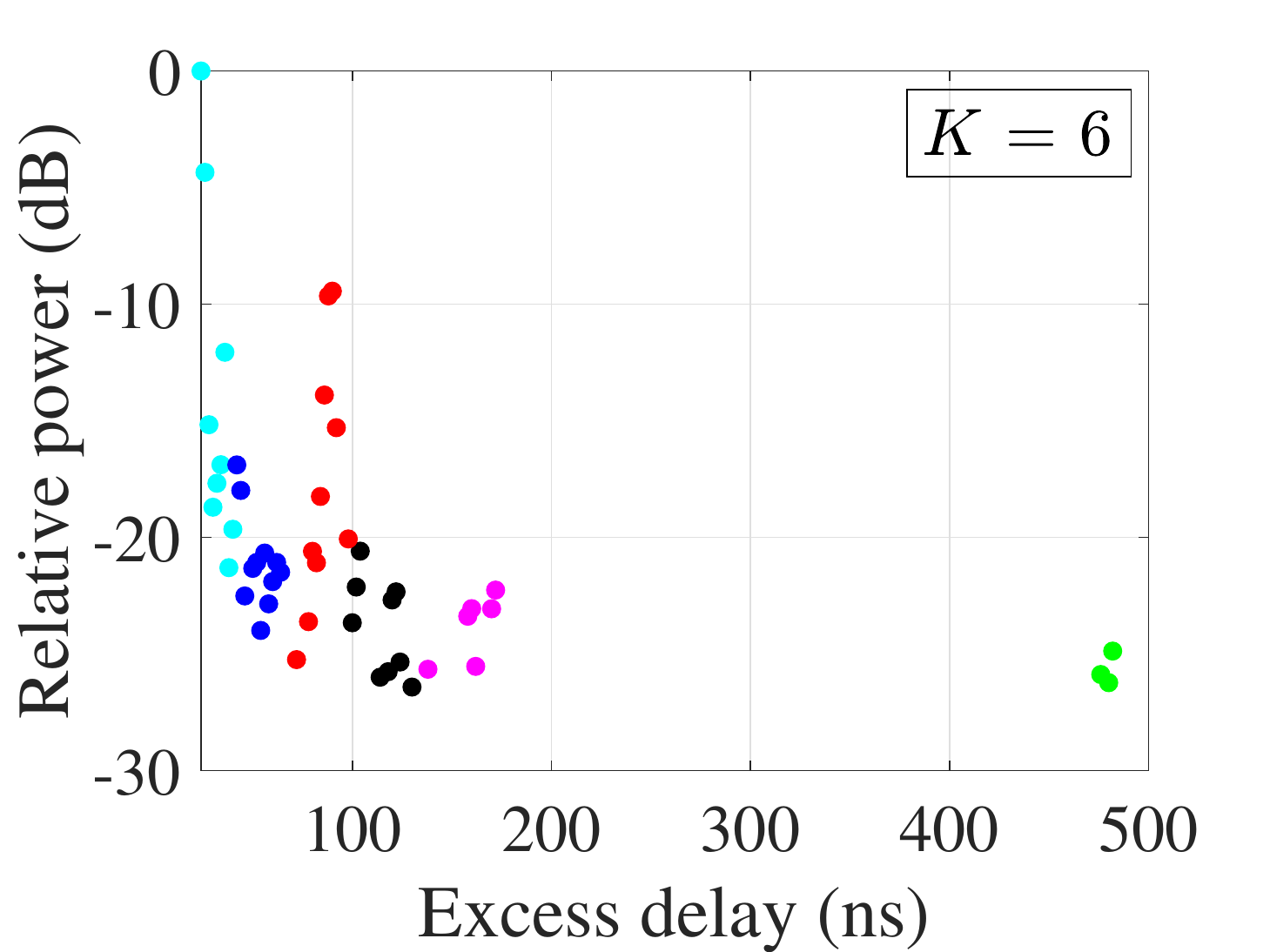}}
   \subfigure[]{\includegraphics[width=1.6in]{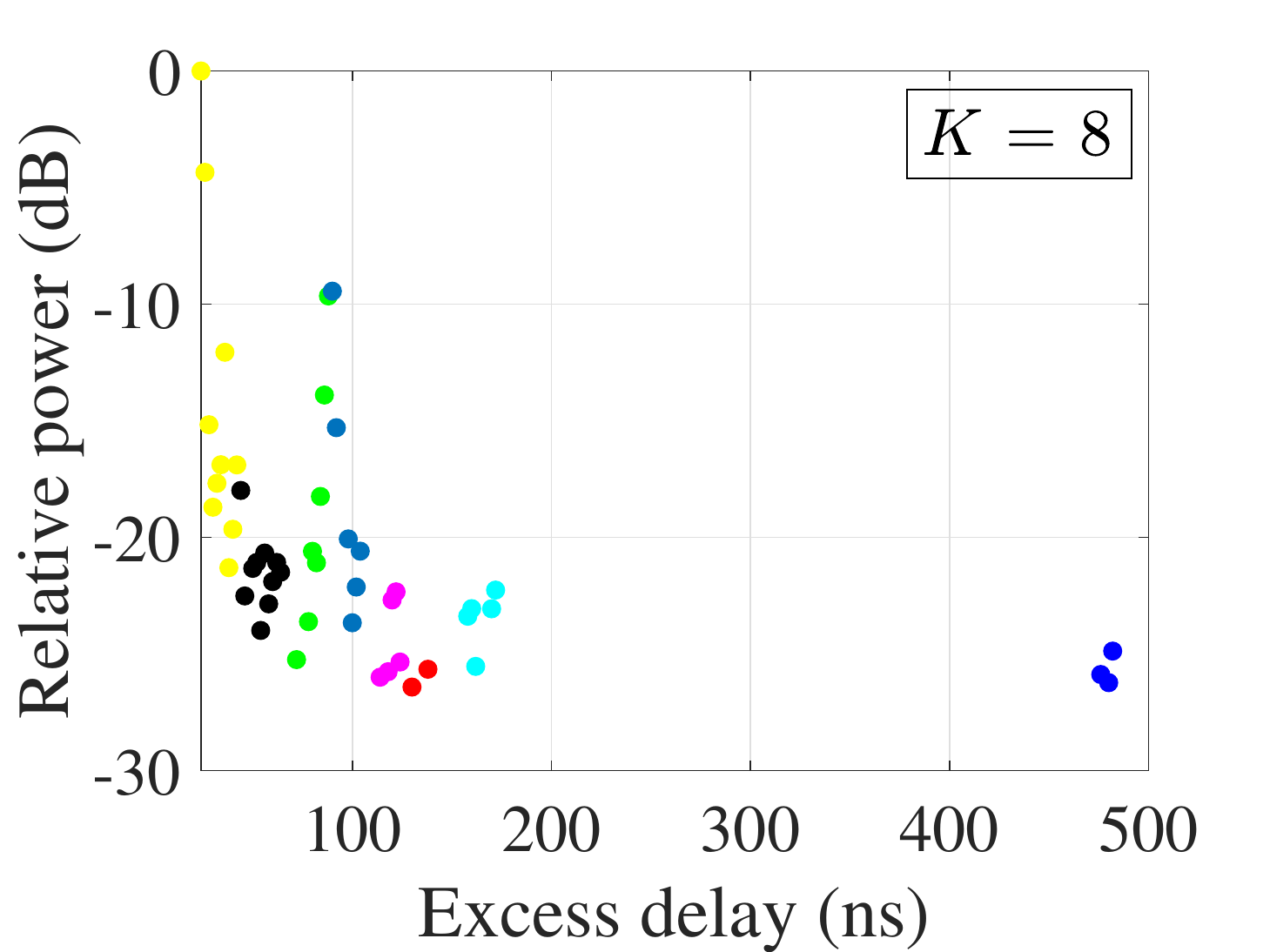}}
    \subfigure[]{\includegraphics[width=1.6in]{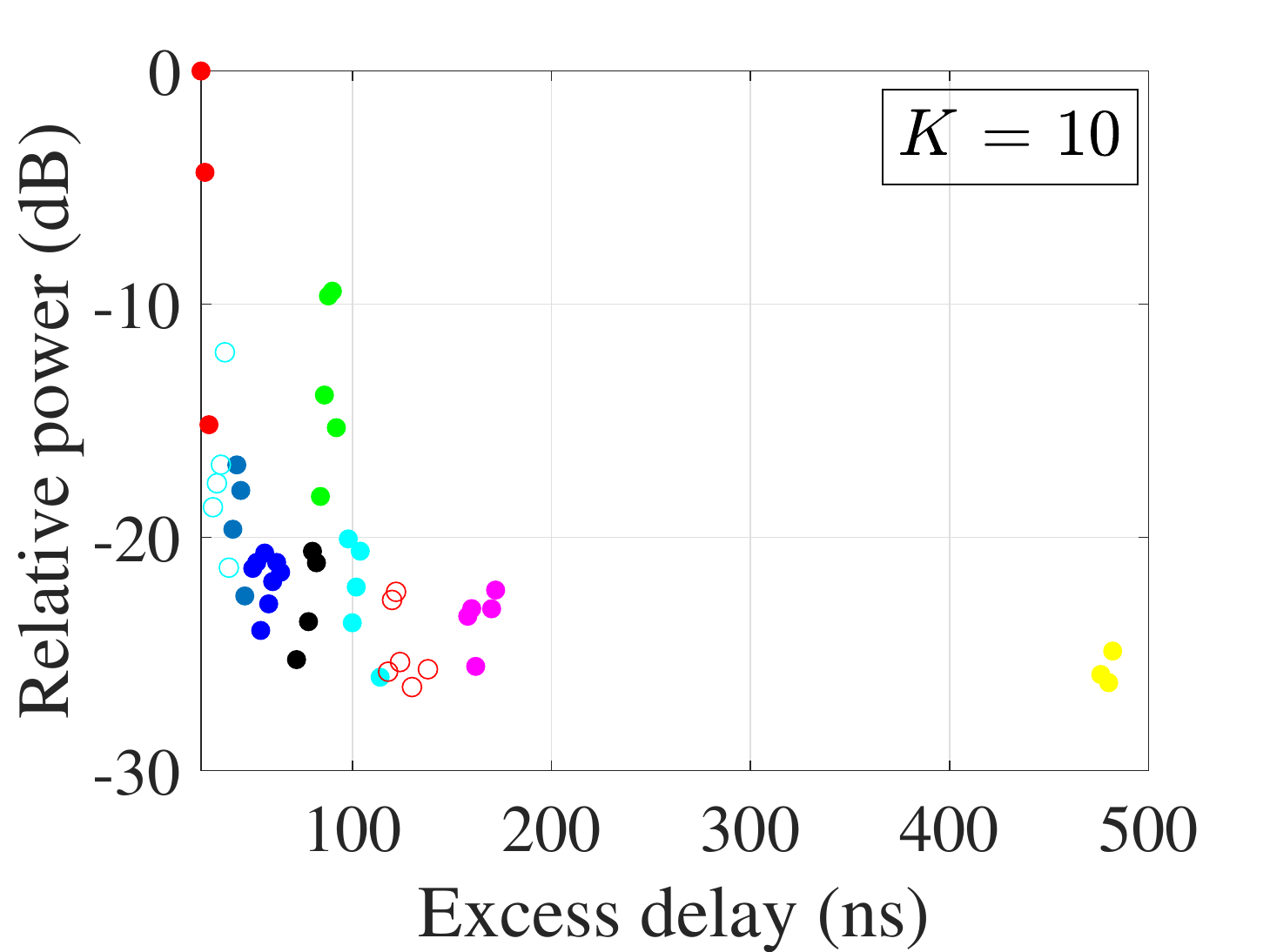}}
  \caption{Clustering  trials: (a)  $K$=4, (b) $K$=6, (c) $K$=8, and (d) $K$=10.}
  \label{kpmtrail}
 \end{figure}

First of all, a reasonable range of cluster numbers can be obtained by clustering trials. As shown in Fig.~\ref{kpmtrail}, we illustrate the clustering results based on KPM with different predefined $K$. It is clearly shown that for $K=4$ and $K=10$, the results lead to under-clustering and over-clustering, respectively. It can be deduced that $[K_{\min}, K_{\max}]$ can be $[4, 10]$ for the collected data in the paper. In particular, the clustering under $K=6$ is more reasonable based on visual observation. For more physical validations, we can first revisit the measurement environment. As shown in Fig.~1, we can observe several scatterer groups in the visible region, which may be sources of cluster formation. According to five possible groups, we can induce that the number of clusters may range from 5 to 7, considering other potential clusters.

 \begin{figure}[!t]
  \centering
 \includegraphics[width=2.8in]{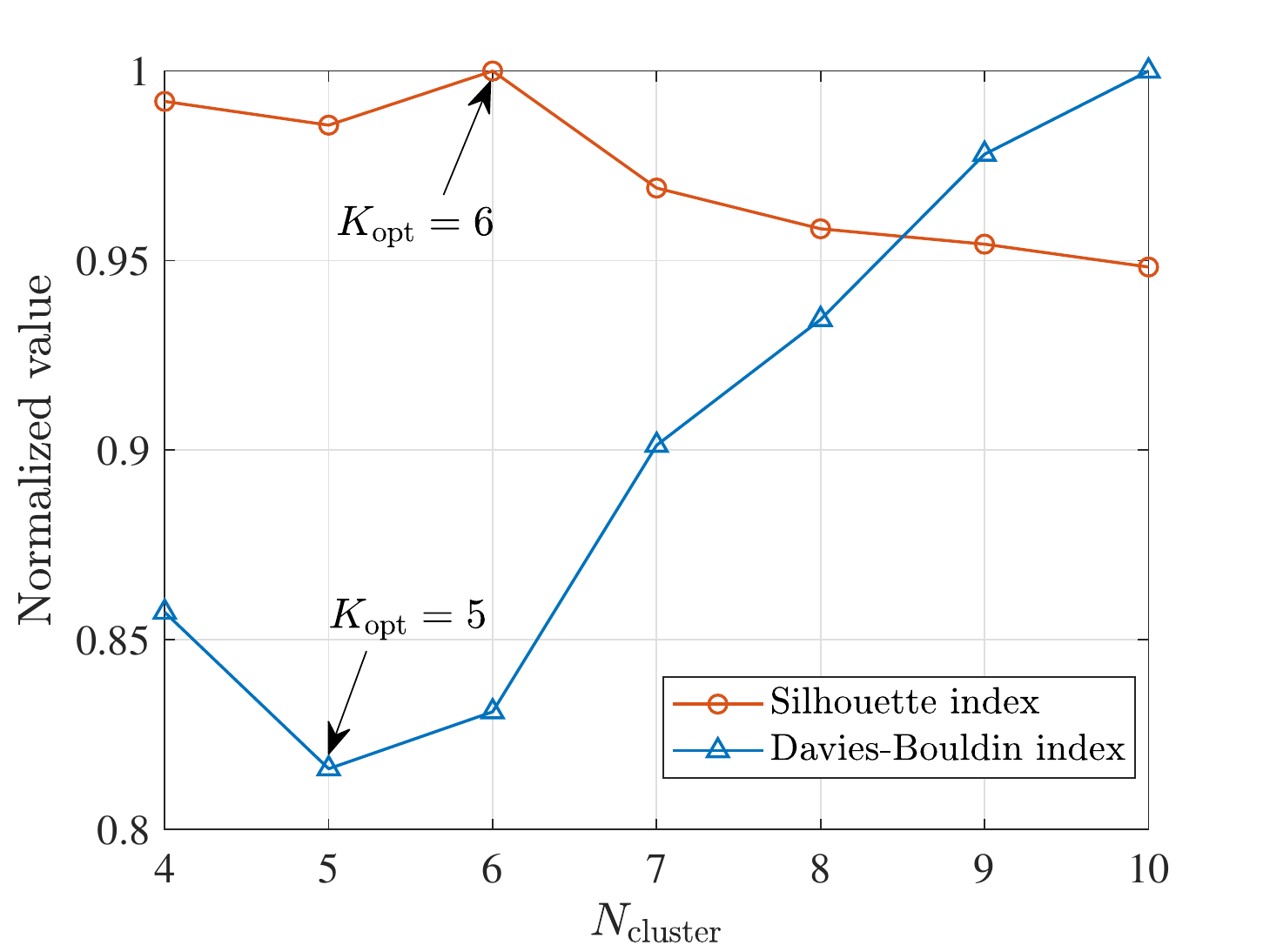}
  \caption{Optimal number of clusters with different evaluation methods.}
  \label{vnc}
 \end{figure}

Mathematically, we then calculate the DB index for different $K$ clusters ranging from 4 to 10, compared to another evaluation method. The results in Fig.~\ref{vnc} suggest that the optimal number of clusters is 5 by employing the DB index, while it becomes 6 under the Silhouette index. Both validations confirm our deduction from the physical perspective. Accordingly, for the whole snapshots, it can be found that the average optimal number of clusters is 5.19 and 6.61 obtained by the DB and Silhouette methods, respectively.
\subsection{Intra-Cluster Characterization} 
\subsubsection{Rectangle Characterization} 
\begin{figure}[!t]
  \centering
 \includegraphics[width=2.8in]{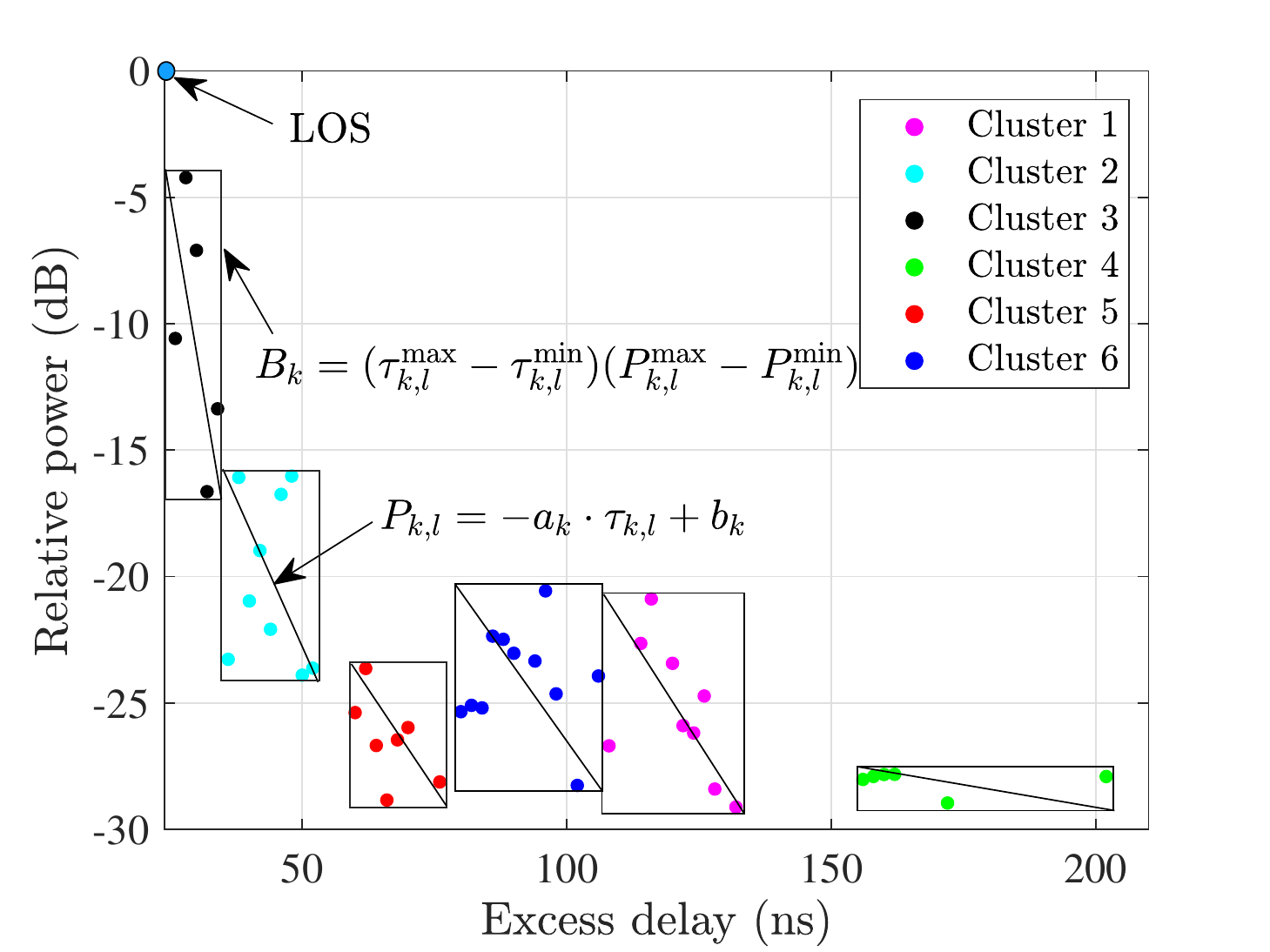}
  \caption{An illustration of the rectangle characterization method.} 
  \label{juxing}
 \end{figure}

For the quantified representation of intra-cluster characteristics, we propose a heuristic characterization method. As shown in Fig.~\ref{juxing}, we introduce rectangles to geometrically characterize clusters, where the rectangle can be determined by four parameters, i.e., $\tau_{k,l}^{\max}$, $\tau_{k,l}^{\min}$, $P_{k,l}^{\max}$, and $P_{k,l}^{\min}$, corresponding to the maximum and minimum delay and power of the $k$-th cluster. Note that we exclude the LOS path in the following characterization and modeling, for the following reasons:~1)~The LOS path presents a large power difference from other MPCs, which is not conducive to clustering performance.~2)~The LOS path can be subsequently incorporated into the channel model according to the real channel state.


 \begin{figure}[!t]
	\centering
	\subfigure[]{\includegraphics[width=1.6in]{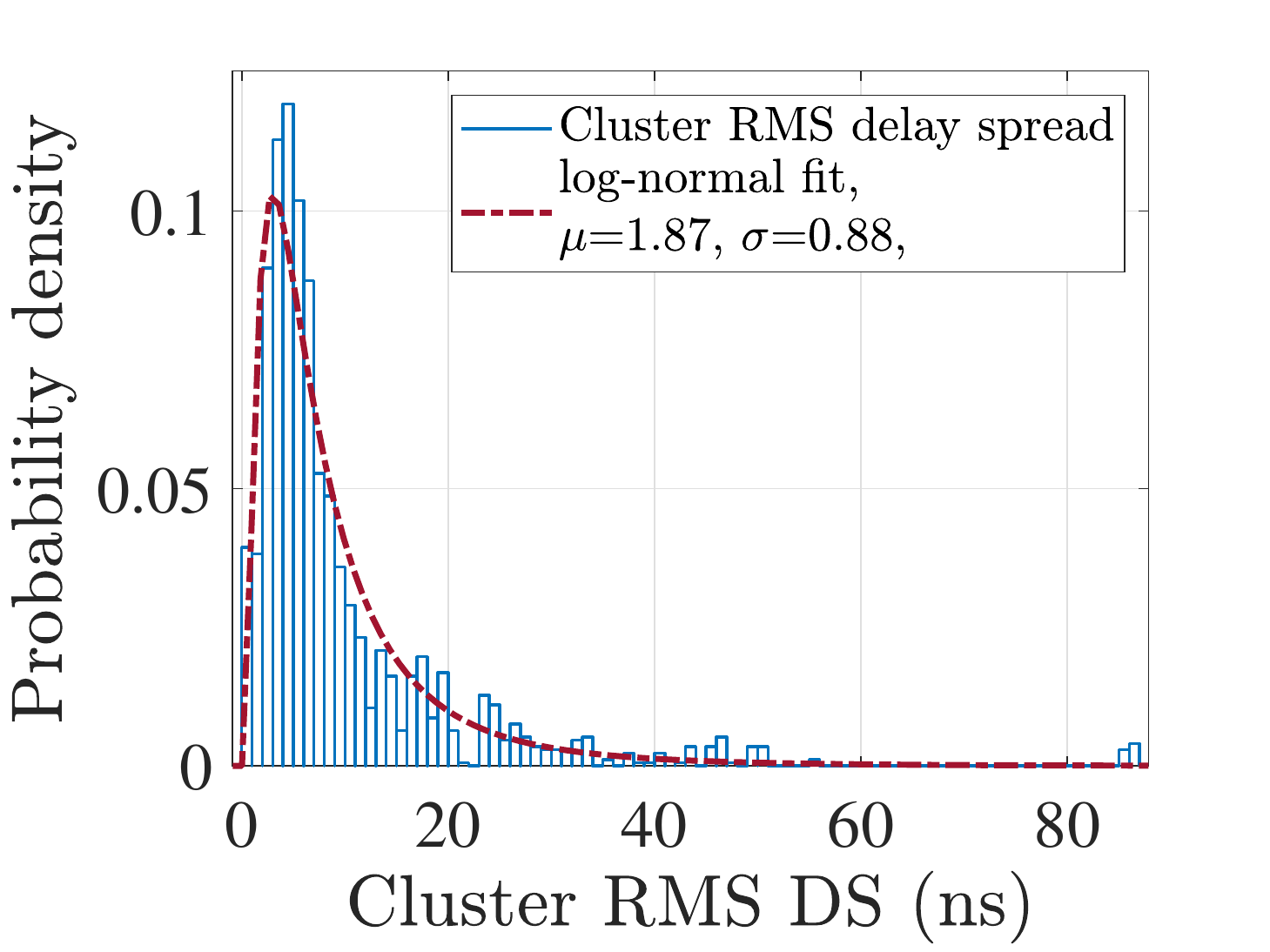}}
	\subfigure[]{\includegraphics[width=1.6in]{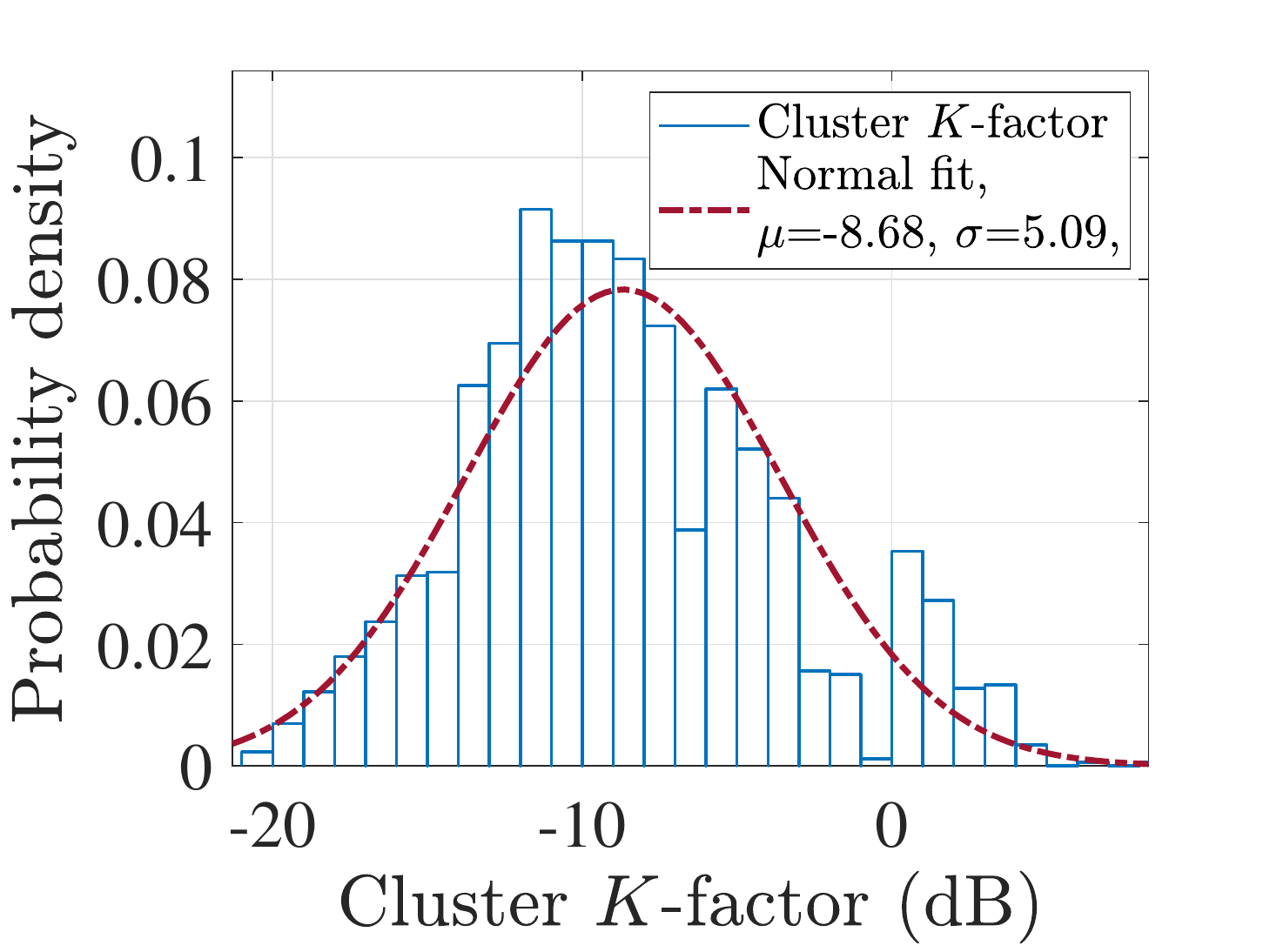}}
	\caption{Intra-cluster parameters: (a) RMS delay spread, (b) Rician $K$-factor.}
	\label{Intra-cluster}
\end{figure}

According to the SV model, the mean power of $l$-th ray in the $k$-th cluster is given by an exponential form \cite{sv}. As a matter of fact, for decibel power, the relationship between power and delay can be represented as a linear form, which can be expressed as
\begin{equation}
P_{k,l}= -\frac{P_{k,l}^{\max}- P_{k,l}^{\min}}{\tau_{k,l}^{\max}-\tau_{k,l}^{\min}}\tau_{k,l}+b_k,
\end{equation}
where the slope ($a_k=(P_{k,l}^{\max}- P_{k,l}^{\min})/(\tau_{k,l}^{\max}-\tau_{k,l}^{\min})$) and intercept $b_k$ represent the power decaying degree and the arrival rate of a cluster, respectively.

In particular, we conduct the linear fitting for all the clusters and thus obtain more than 2000 sets of $a_k$. We found that it follows the Weibull distribution $\mathcal{W}(0.55, 1.21)$. Moreover, the mean value $\overline{a_k}$ is 0.53 dB/ns, showing the power decay degree of intra-cluster on average.

For a better characterization, we determine the ray unit area $A_k$ in the cluster whose area is $B_k$, which can be expressed as
\begin{equation}
A_{k}= B_k/L_k=(\tau_{k,l}^{\max}-\tau_{k,l}^{\min})\cdot(P_{k,l}^{\max}- P_{k,l}^{\min})/L_k,
\end{equation}
where $A_{k}$ can be used for measuring the contribution of the ray to the cluster. It is found that $A_{k}$ follows Weibull distribution $\mathcal{W}(25.75, 1.46)$. Moreover, it is important for the cluster-based model to determine the reasonable range of generated delay and power. For instance, for a given $L_k$, we can determine the range of delay with $\tau_{k,l}^{\max}-\tau_{k,l}^{\min}=\sqrt{(A_k L_k)/a_k}$.


\begin{figure}[!t]
	\centering
	\includegraphics[width=2.8in]{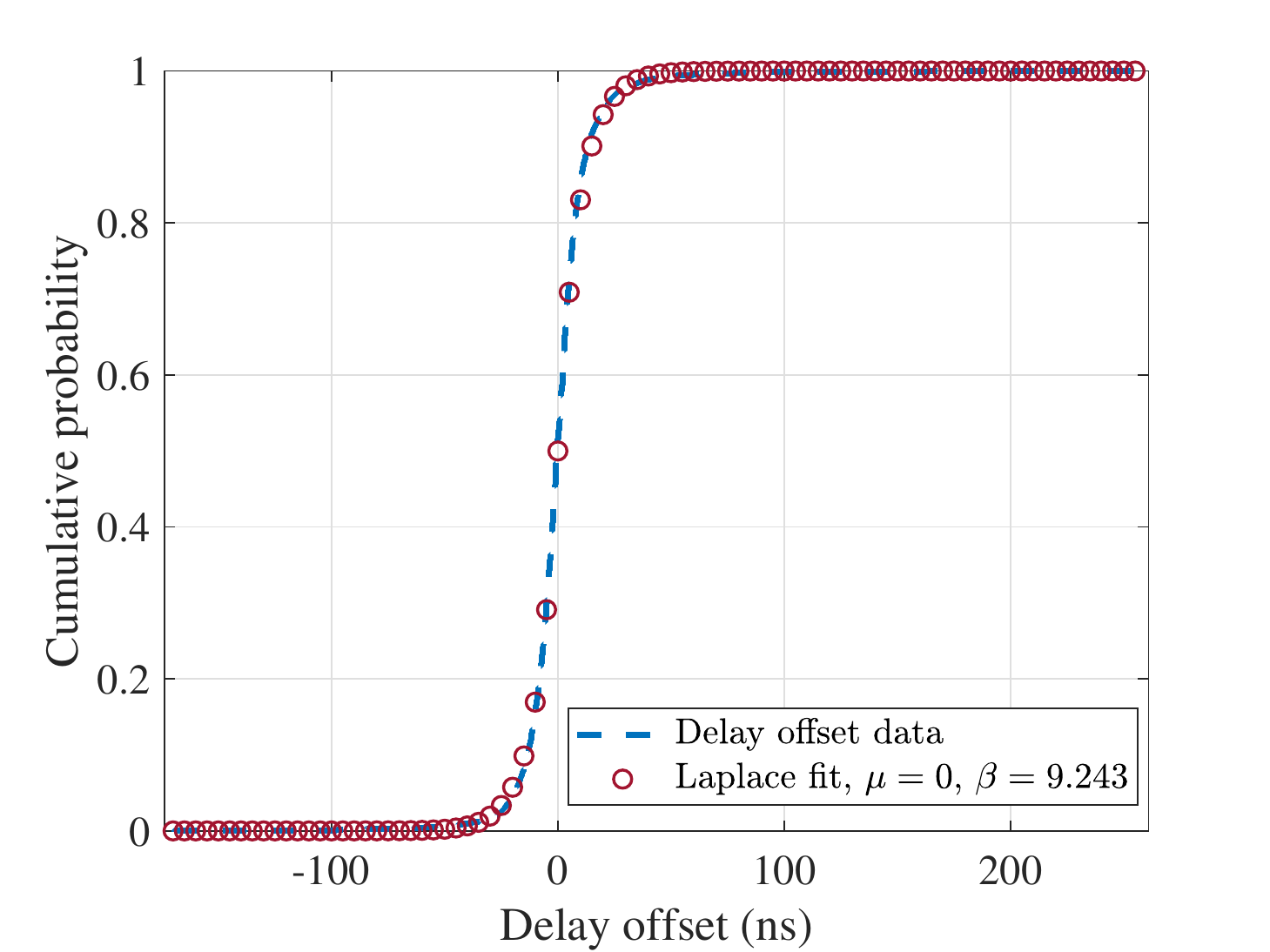}
	\caption{CDF of delay offset with Laplace fitting $\mathcal{L}(0, 9.243)$.}
	\label{offset}
\end{figure}

\subsubsection{Cluster Delay Spread and Rician $K$-Factor} Rician $K$-factor (KF) and root-mean-square (RMS) delay spread for intra-cluster are investigated herein. The cluster $K$-factor is defined as the power ratio between the strongest MPC and the summation of remaining MPCs in the cluster, given by
\begin{equation}
KF_{k} [{\rm dB}]= 10\log_{10}\frac{\max\limits_l{(a_{k,l}^2)}}{\sum_l^{L_k}{(a_{k,l}^2)}-\max\limits_l{(a_{k,l}^2)}}.
\end{equation}
The RMS delay spread describes the dispersion of multipath channels, defined as the square root of the second central
moment of PDPs. Specifically, the cluster RMS delay spread can be calculated by
\begin{equation}
\sigma_{\tau_k} = \sqrt{\frac{\sum_{l=1}^{L_k}(\tau_{k, l}-\bar{\tau})^2 a_{k, l}^2}{\sum_{l=1}^{L_k} a_{k, l}^2}},
\end{equation}
where the power-weighted average delay is
\begin{equation}
\label{eq:mean excess delay}
 \bar{\tau}=\frac{\sum_{l=1}^{L_k}\tau_{k, l} a_{k, l}^2}{\sum_{l=1}^{L_k} a_{k, l}^2}.
\end{equation}

We show the probability density functions (PDFs) of the cluster RMS delay spread and $K$-factor in Fig.~\ref{Intra-cluster}.  Notably, all the PDFs and cumulative distribution functions (CDFs) are carefully checked by the Kolmogorov-Smirnov (KS) test and shown to be the best fit among popular distributions such as Normal, $\log$-normal, Rician, Rayleigh, Weibull, and Exponential distributions, etc. Subsequent fits in the sequel are also verified in the same way. It is found that the cluster RMS delay spread follows the $\log$-normal distribution, which is expressed as $\ln{\sigma_{\tau_c}}\sim \mathcal{N}(1.87, 0.88)$. In addition, cluster $K$-factor follows the Gaussian distribution $K_c \sim \mathcal{N}(-8.68, 5.09)$. The small $K$-factors further confirm that the MPC power in a cluster presents the Rayleigh distributed characteristic, which can be well explained by the nature of clustering that concentrates on power and delay similarities, without a dominant component. Worth noting that the corresponding parameters are summarized in Table~I.

    \begin{figure}[!t]
  \centering
  \subfigure[]{\includegraphics[width=3in]{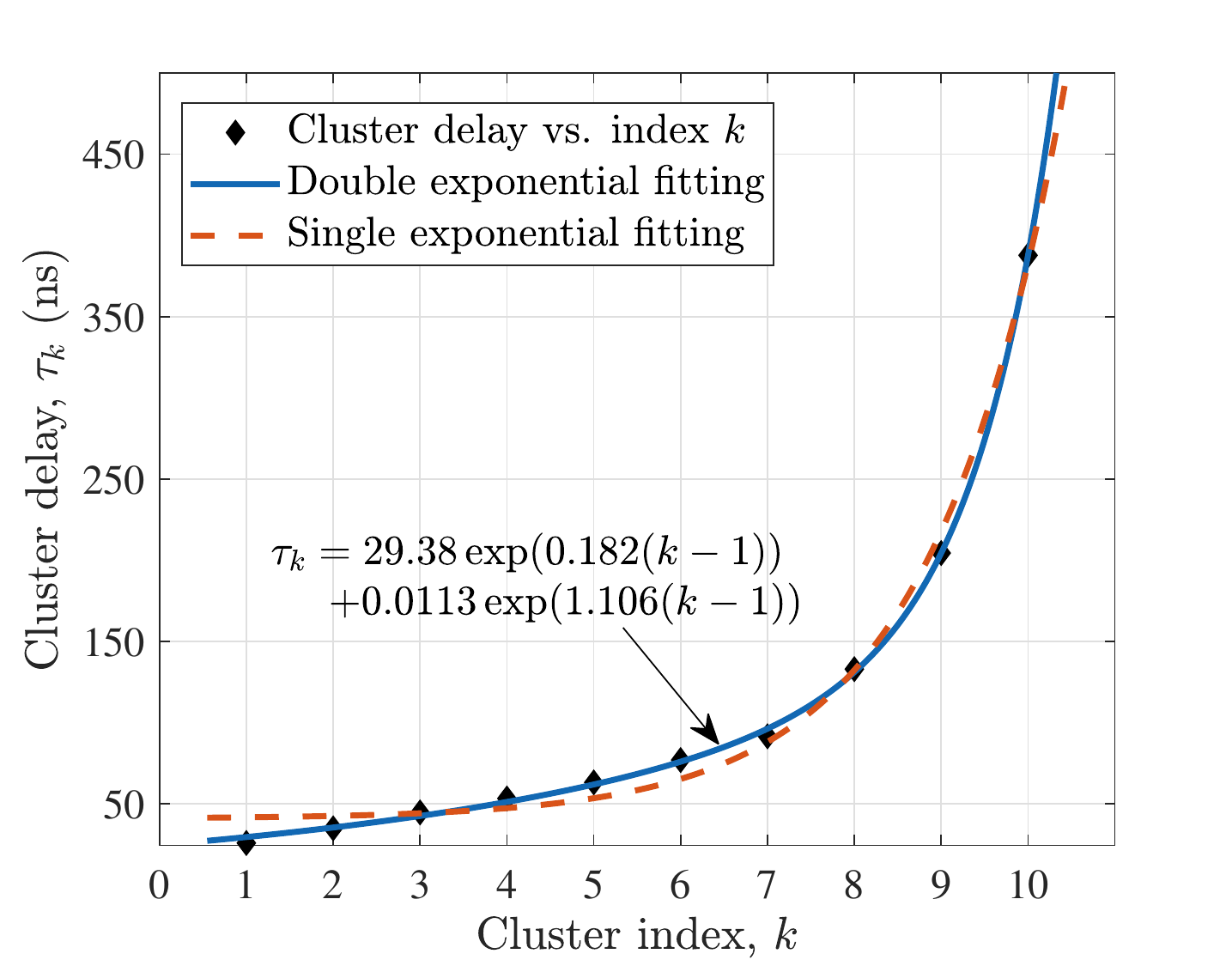}}
  \subfigure[]{\includegraphics[width=3in]{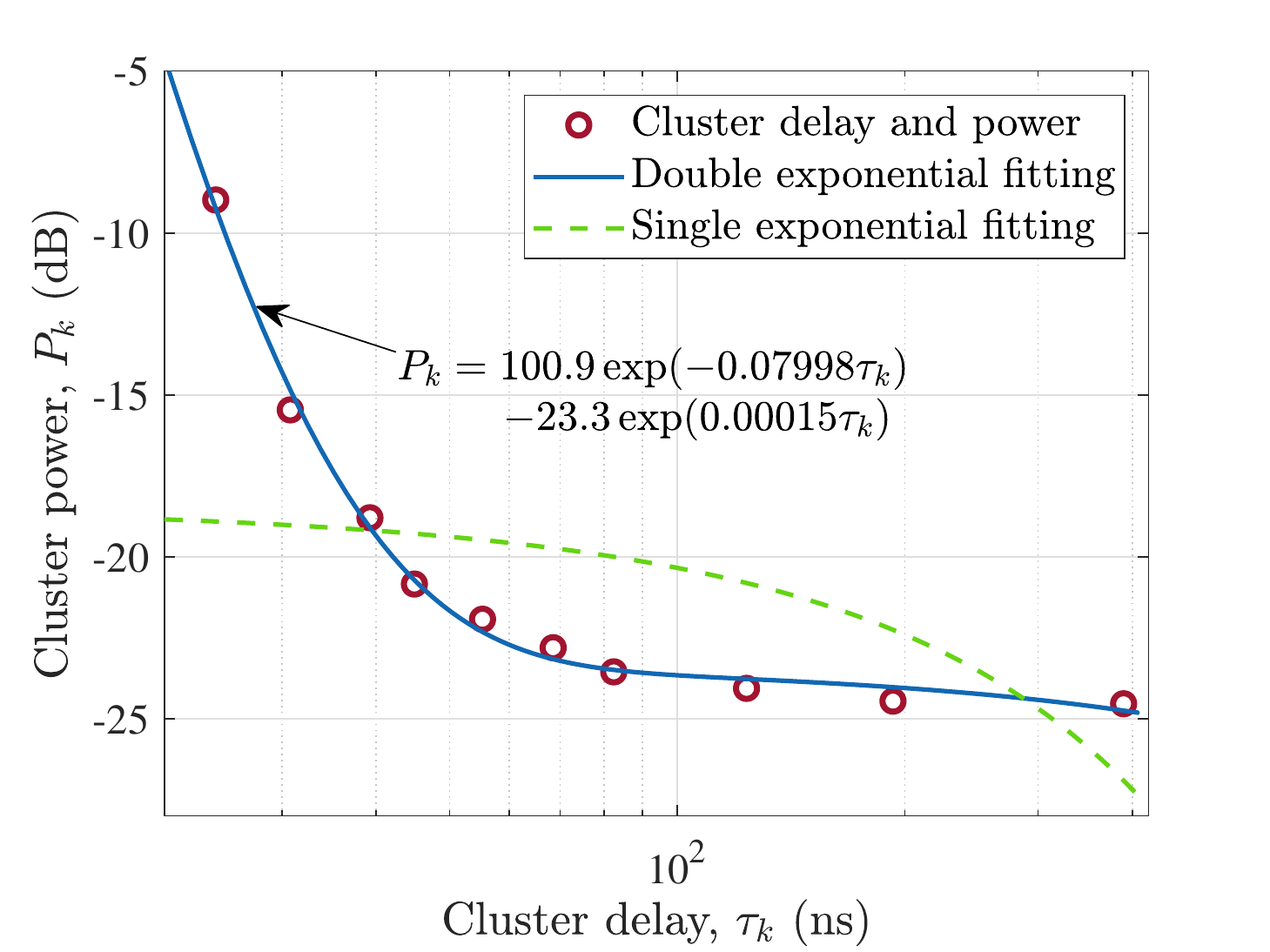}}
  \caption{Inter-cluster modeling: (a) delay vs. index, (b) power vs. delay.}
  \label{clu_dp}
 \end{figure}

\subsubsection{Intra-Cluster Delay Offset} For generating the MPC delays in a cluster, it is necessary to characterize the intra-cluster delay offset. We denote the delay offset as $\tau_{\rm os}$ that can be calculated by $\tau_{k,l}-\overline{\tau_{k,l}}$, where $\overline{\tau_{k,l}}$ represents the cluster delay that is an average of all MPC delays in a cluster. As shown in Fig.~\ref{offset}, we show the CDF of $\tau_{\rm os}$ and corresponding fit. It is found that the Laplace distribution can perfectly match the empirical data. In fact, the Laplace distribution is widely used in describing both the delay and angle offsets in cluster-based channel model \cite{jzl}. The obtained parameter can be used for generating MPC delays in the stochastic channel model.

\subsection{Inter-Cluster Characteristics}
The inter-cluster characterizations mainly include the number of clusters, the birth-and-death characteristics of clusters, the power decay and delay modeling of inter-clusters, and the occurrence probability of clusters. We have investigated the optimal number of clusters through two different test methods in the prior section. Therefore, we herein will focus on the remaining parameters, highlighting the tracking process.

 \begin{table}[t]
	\centering
	\caption{Cluster-based Model Parameters} \label{table1}
	\begin{tabular}{l|l|l}
		\hline
		Parameter &  Notation & Value \\
		\hline
		\multicolumn{3}{c}{Inter-Cluster Characteristics}    \\
		\hline
		\multirow{2}*{Cluster number} & \multirow{2}*{$N_{\text{c}}$}\;\;\; Silhouette & $\mu_s$=6.61, $\sigma_s$=2.07\\
		~ & ~ \;\;\;\;\;\;\;\; DB & $\mu_d$=5.19, $\sigma_d$=1.46 \\
		
		Cluster survival length  & $S_{\rm CL}$ [m]   & $p$=7.11, $q$=1.47\\
		
		\multirow{2}*{Occurrence probability}  & \multirow{2}*{$P_{\rm oc}$} \; $k \le 4$   &  \;\;\;\;\;\;\;\;\;\;\;1 \\
		~ & ~ \;\;\;\;\;  $4<k \le 10$ & -0.115$k$+1.361\\
		\hline
		\multicolumn{3}{c}{Intra-Cluster Characteristics}    \\
		\hline
		\multirow{2}*{Ray number per cluster} & \multirow{2}*{$L_{\text{ray}}$}& 7.41  (Silhouette)\\
		~ & ~ & 9.44\;\;\;\; (DB) \\
		Ray unit area & $A_k$ (Weibull) [dB$\cdot$ns] & $p$=25.75, $q$=1.46 \\
		Cluster $K$-factor  & $KF_k$ [dB]   & $\mu_{\rm K}$=-8.68, $\sigma_{\rm K}$=5.09 \\
		Cluster RMS DS & $\sigma_{\tau_k}$ [ns]   & $\mu_{\text{DS}}$=1.87, $\sigma_{\text{DS}}$=0.88 \\
		\multirow{1}*{Intra-power decay}& $a$ (Weibull)  [dB/ns] & $p$=0.55, $q$=1.21  \\
		
		Delay offset& $\tau_{\rm os}$ (Laplace) [ns]  & $\mu$=0, $\beta$=9.243 \\
		\hline
	\end{tabular}
\end{table}

\subsubsection{Power and Delay Modeling of Inter-Cluster}
Based on the clustering results, we obtain the average delay and power of clusters, denoted as $\tau_k$ and $P_k$, respectively, where $\tau_k=\overline{\tau_{k,l}}$ and $P_k=\overline{P_{k,l}}$. We first model the relationship between the cluster delay and its index, which facilitates generating delay with the given number of clusters. As shown in Fig.~\ref{clu_dp}(a), it is found that a single exponential fitting fails to describe the delay-index relation for a small index. Thus, we employ a double exponential fitting that is capable of well describing the relationship, which can be expressed as 
\begin{equation}
	\tau_k=29.38\exp(0.183(k-1))+0.0113\exp(1.106(k-1)).
\end{equation}

With the empirical result, we can generate cluster delay for a given number of clusters. Then, we illustrate the relationship between cluster delay and power, as shown in Fig.~\ref{clu_dp}(b). In the SV model, a single exponential form represents the relation. However, we found that the single exponential fitting of cluster delay and power appears a vast divergence from the actual situation. Thus, we employ a double exponential fitting, which results in a considerable agreement with empirical data. Specifically, the expression is given by 
\begin{equation}
	P_k=100.9\exp(-0.07998\tau_k)-23.3\exp(0.00015\tau_k).
\end{equation}

With generated cluster delay, we can determine the cluster power with this expression. Subsequently, the intercept $b_k$ can be determined by $b_k=P_k+a_k\tau_k$, which can be further used to generate the power of sub-path in a cluster.
\subsubsection{Clustering-based Tracking Method} In the prior tracking method, a series of successive clusters with intervals of dozens of wavelengths make up a trajectory \cite{ch2}. However, the trajectory will be interrupted when the distance between neighboring clusters is beyond the given threshold. Accordingly, a new trajectory needs to be shaped. This method focuses on the short-term tracking that only considers adjacent clusters in a very short distance, which may lead to over-tracking and need too many trajectories. To compensate for the lack, we consider that distant clusters can also constitute a trajectory if they have similar delay and power. In this regard, we develop a long-term tracking method, taking the joint delay and power similarities of clusters into consideration. 

We denote $A_{n,1}$, $A_{n,2}$, ..., $A_{n, c_n}$ and $A_{m,1}$, $A_{m,2}$, ..., $A_{m, c_{m}}$as clusters in $n$- and $m$-th snapshots, respectively. $c_n$ and $c_{m}$ represent the number of clusters, which are determined by the DB test. Specifically, we aim to associate all \emph{similar} clusters for the whole snapshots. Notably, the difference between our proposed method and traditional tracking algorithm lies in that the subscript $m$ can be any snapshot in our method, while it merely can be $n+1$ or $n-1$ in previous works \cite{ch1,zyh}.

As for the measure of similarity, we use the weighted 3D Euclidean distance, denoted as $\mathcal{D}$, which is given by
\begin{equation}
\begin{aligned}
&\mathcal{D}~(A_{n,c_n}, A_{m,c_{m}}) \\& =\sqrt{w_d(d_{n}-d_m)^2+w_p(p_{n,c_n}-p_{m,c_m})^2+w_{\tau}(\tau_{n,c_n}-\tau_{m,c_m})^2},
\end{aligned}
\end{equation}
where $p_{n,c_n}$ and $\tau_{n,c_n}$represent the normalized power and delay of the $c_n$-th cluster in the $n$-th snapshot, respectively, which are obtained by averaging the power and delay of all MPCs in a cluster. $d_{n}$ is the link distance between Tx and Rx in the $n$-th snapshot. Moreover, $w_d$, $w_p$ and $w_{\tau}$ are the weights of link distance, power and delay, respectively. We then employ a clustering-based tracking (CBT) approach to trace the trajectories of clusters. 

  \begin{figure}[!t]
	\centering
	\subfigure[]{\includegraphics[width=1.6in]{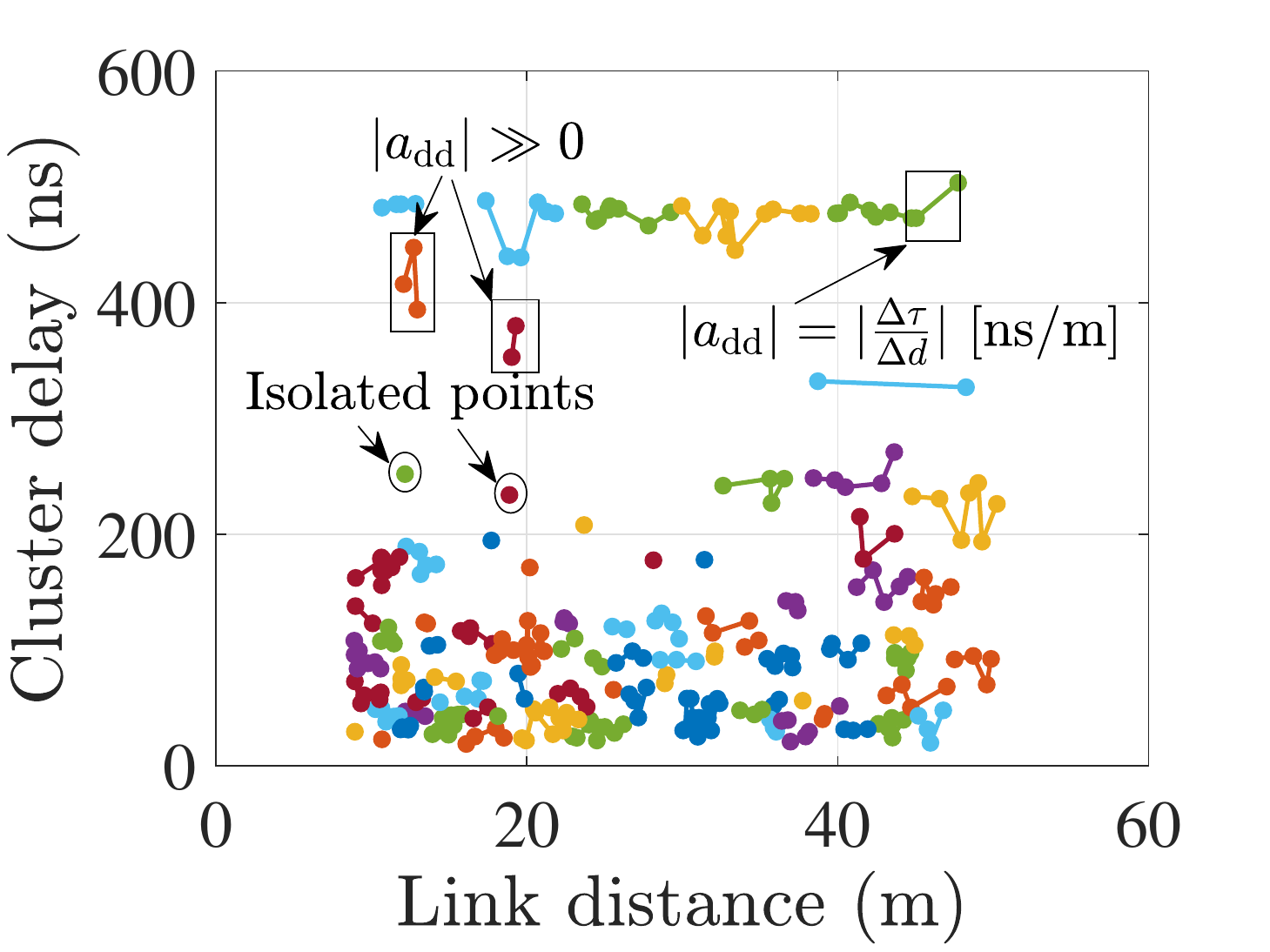}}
	\subfigure[]{\includegraphics[width=1.6in]{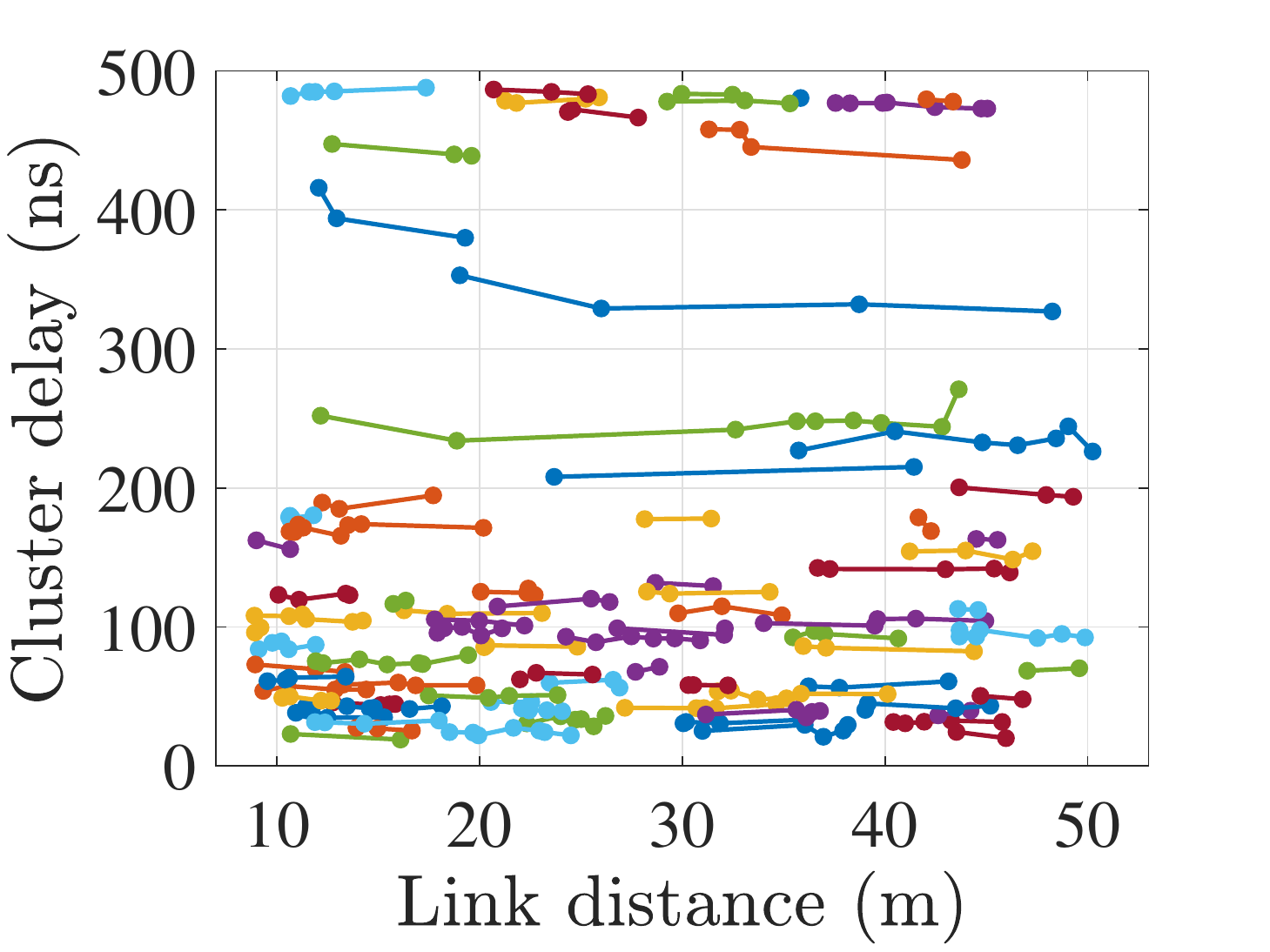}}
	\subfigure[]{\includegraphics[width=1.6in]{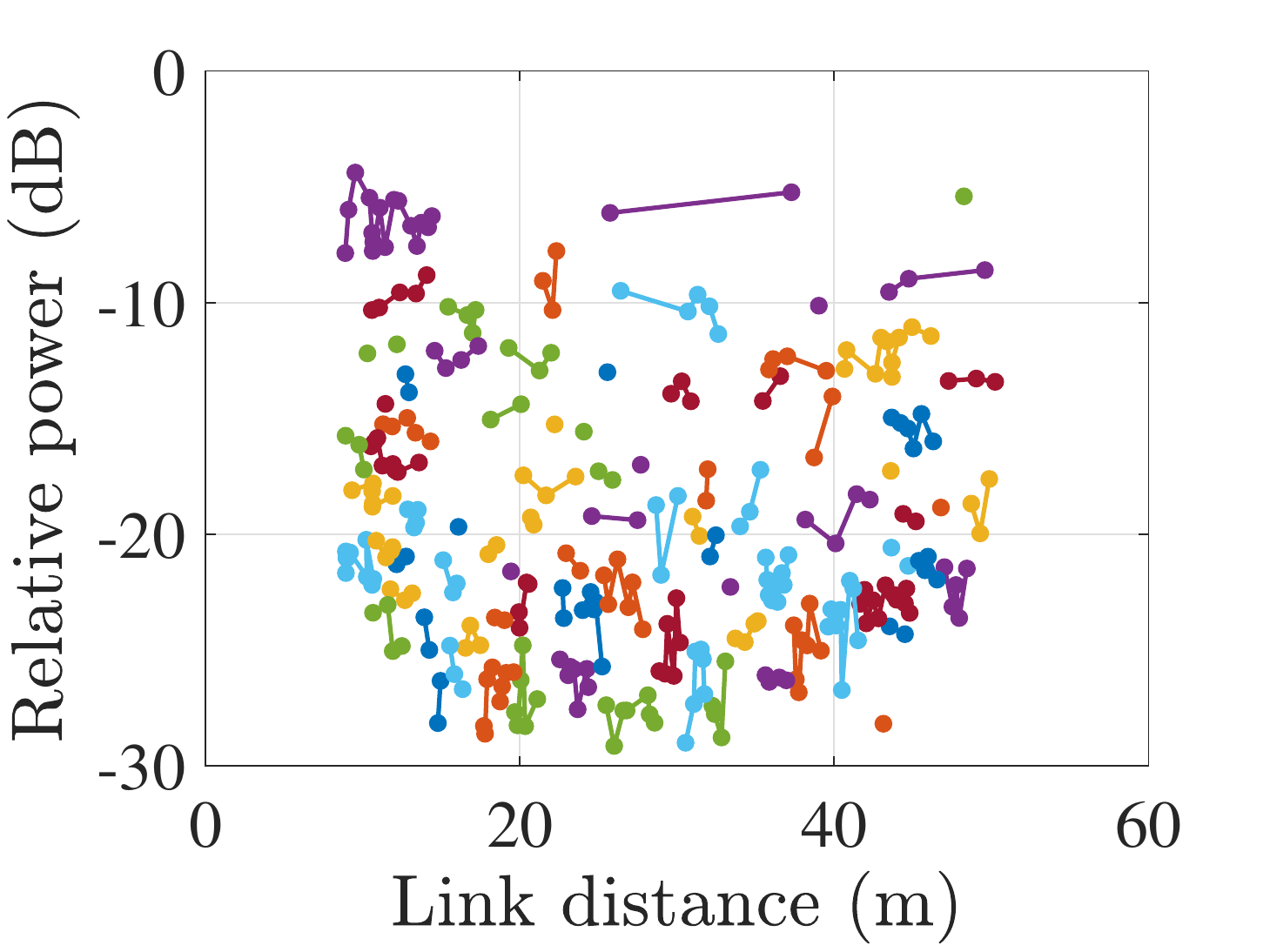}}
	\subfigure[]{\includegraphics[width=1.6in]{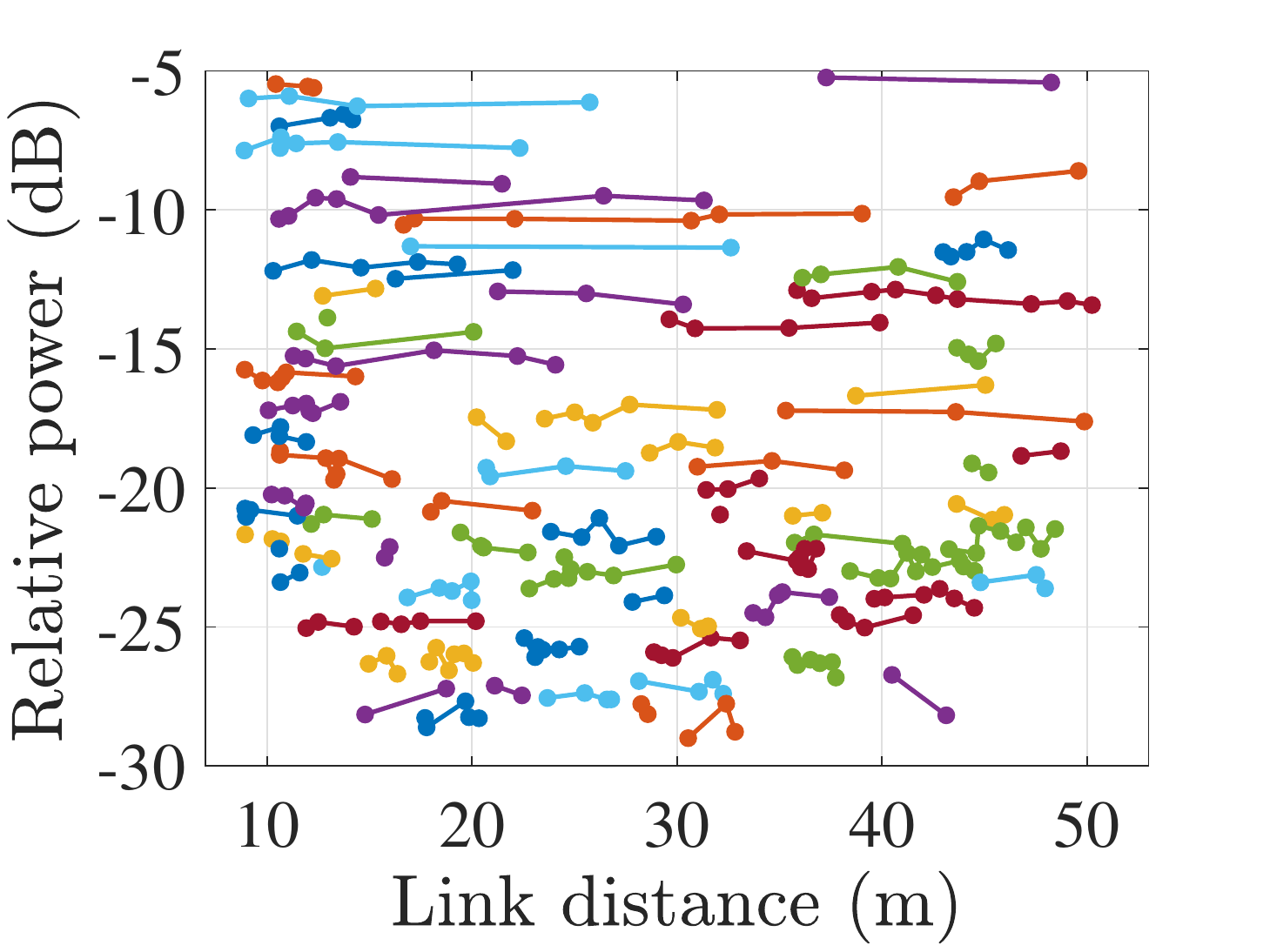}}
	\caption{2D tracking results: (a) unweighted distance/delay, (b) weighted distance/delay, (c) unweighted distance/power, and (d) weighted distance/power.}
	\label{trackingddp}
\end{figure}
\begin{figure}[!t]
  \centering
{\includegraphics[width=2.8in]{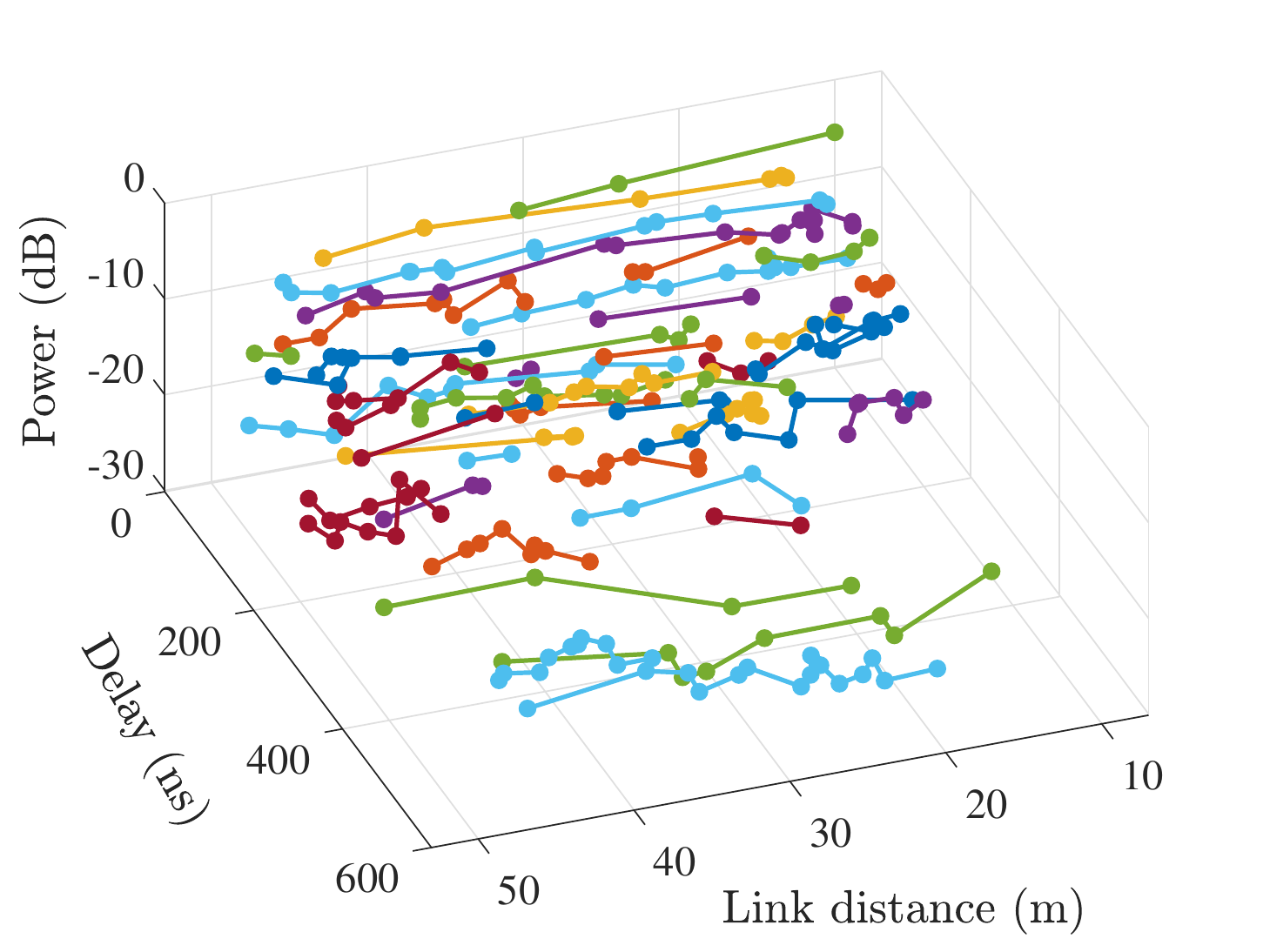}}
  \caption{Cluster tracking results under the weighted 3D Euclidean distance.}
  \label{tracking}
 \end{figure}

Our objective is to track the clusters with similarity in delay or power and the continuity in the distance. Accordingly, we first show the 2D tracking results in Fig.~\ref{trackingddp}, where (a) and (b) consider the unweighted and weighted distance-delay measure, whereas (c) and (d) consider the unweighted and weighted distance-power measure. For unweighted measures, we employ $w_d=w_p=w_{\tau}=1$, which improperly take the distance similarity into consideration. As shown in Fig.~\ref{trackingddp}(a), we illustrate the slope of trajectory for a better understanding, where $a_{\rm dd}=\frac{\Delta \tau}{\Delta d}$. It confirms that the values of $a_{\rm dd}$ should be close to zero. As an example, we use the ground reflection path to justify our findings. Without loss of generality, we denote the path length of reflection and LOS as $l_1$/$l_2$ and $d_1$/$d_2$ for two snapshots, respectively. Thus, $a_{\rm dd}$ can be calculated as 
\begin{equation}
	\begin{aligned}
		a_{\rm dd} &=\frac{\Delta \tau}{\Delta d}=\frac{(l_2-l_1)-(d_2-d_1)}{c}\frac{1}{d_2-d_1}\\
		&=\left(\frac{l_2-l_1}{d_2-d_1}-1\right)\frac{1}{c} \approx 3.33 \left(\frac{l_2-l_1}{d_2-d_1}-1\right) \rm{[ns/m]},
	\end{aligned}
\end{equation}
where $c$ is the speed of light. For the path length, $l_{1,2}=\sqrt{(h_a+h_g)^2+D_{1,2}^2}$ and $d_{1,2}=\sqrt{(h_a-h_g)^2+D_{1,2}^2}$, where $D_{1,2}$ is the horizontal distance between Tx and Rx. $h_a$ and $h_g$ are the UAV and ground station heights, respectively. In our measurement, $h_g$ is as low as 0.5~m. Thus, we have $\frac{l_2-l_1}{d_2-d_1} \to 1$, and thus $|a_{\rm dd}| \to 0$ and $|a_{\rm dd}| \neq 0$. Fig.~\ref{trackingddp}(a) shows that there are many slopes much larger than 0 with an unweighted measure. To address this issue, we first identify that the weights can measure the contributions of corresponding variables in the clustering. For example, for $w_d=0$, the trajectory will be formed by clusters with the same delay. It indicates that the condition $w_d=0$ corresponds to $a_{\rm dd}=0$. Moreover, when $w_d=w_{\tau}=1$ (unweighted), the value of $|a_{\rm dd}|$ is close to 1, which is liable to understand because the contributions of distance and delay are identical. Thus, a heuristic relation between the slope and weights can be expressed as $|a_{\rm dd}| \sim |\frac{w_d}{w_{\tau}}|$. Due to small $|a_{\rm dd}|$, we set  $w_d$ and $w_{\tau}$ as 0.05 and 0.95, respectively. The result is illustrated in Fig.~\ref{trackingddp}(b), which shows more reasonable trajectories. The 2D tracking with power and distance considers the same setting, and the results are shown in Fig.~\ref{trackingddp}(d). For the 3D tracking, we consider similarities in power and delay domains and continuity in the link distance domain. Thus, we set $w_d=0.05$, $w_{\tau}=w_p=0.95$. Fig.~\ref{tracking} shows many cluster trajectories with different lengths that well manifest the birth-and-death characteristics of clusters. 

  \begin{figure}[!t]
  \centering
{\includegraphics[width=2.8in]{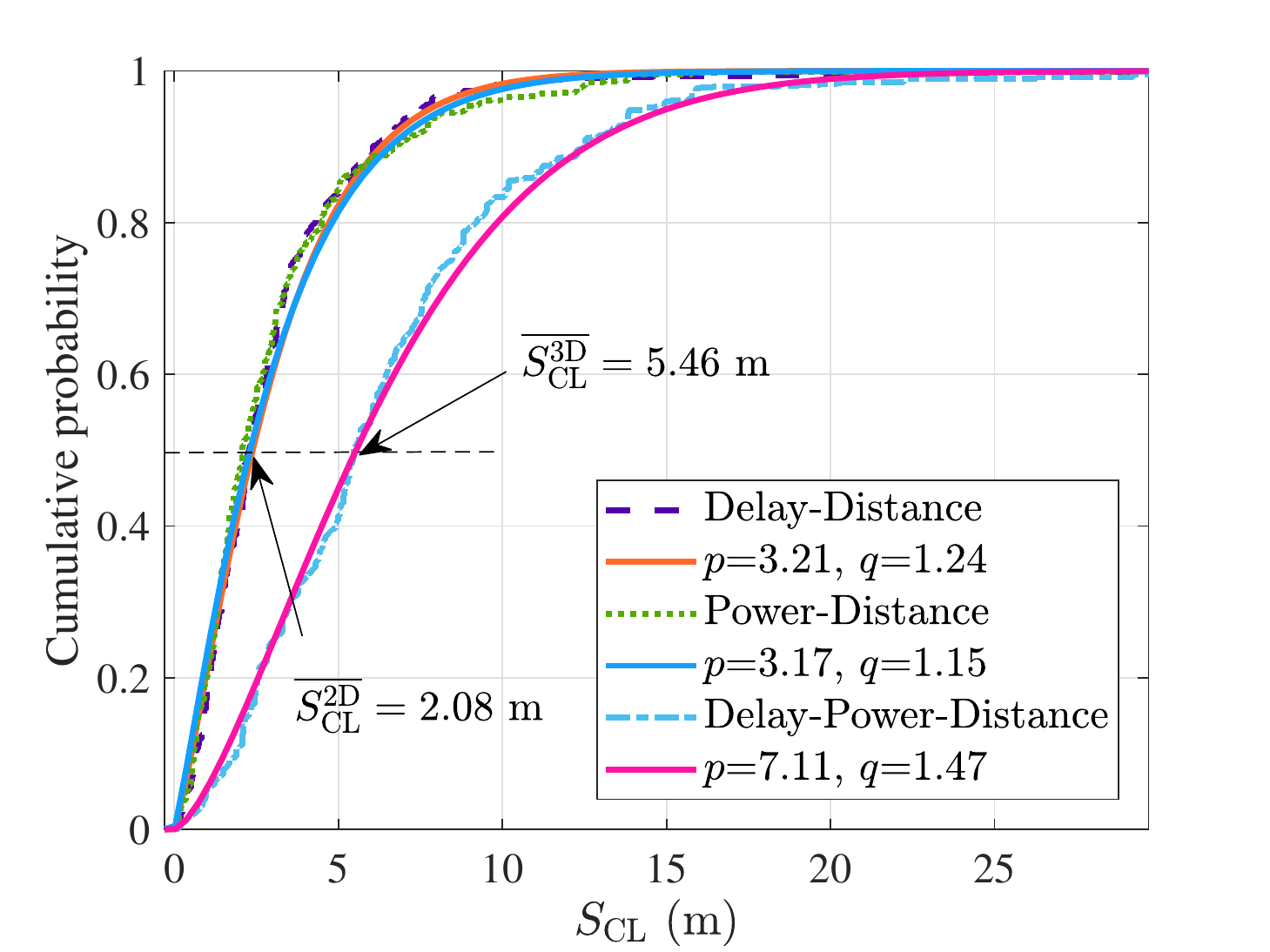}}
  \caption{CDFs and Weibull fits of survival lengths for 2D and 3D tracking.}
  \label{SD}
 \end{figure}

   \begin{figure}[!t]
	\centering
	{\includegraphics[width=2.8in]{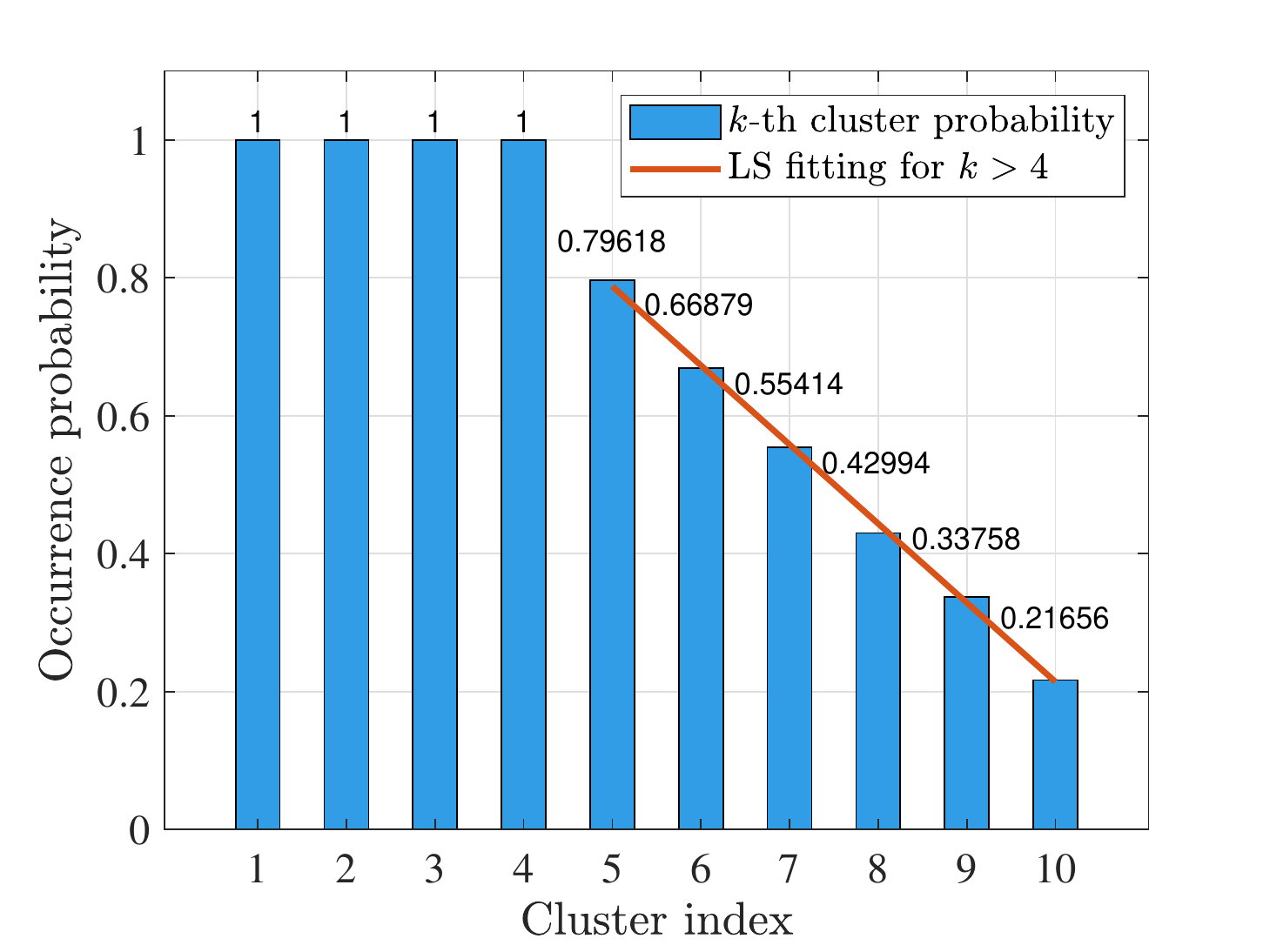}}
	\caption{Histogram of the occurrence probability of the $k$-th cluster.}
	\label{OP}
\end{figure}

\subsubsection{Observed Length of Clusters}
To investigate the dynamic characteristics of clusters, the length of trajectory is a prevalent measure \cite{QW}. It is defined as the continuous distance between the appearance and disappearance of clusters, denoted as $S_{\rm CL}$. For 2D and 3D tracking, the CDFs of observed lengths are shown in Fig~\ref{SD}. We found that the Weibull distribution is capable of describing the survival distances of clusters. For either delay-distance or power-distance tracking, the lengths have similar distributions with the mean $S_{\rm CL}^{\rm 2D}$ of 2.08~m. The trajectory length increases in the 3D tracking where the mean of $S_{\rm CL}^{\rm 3D}$ is 5.46~m. The increase of length can be properly explained by the increase of dimensions in the tracking. Since we solely consider power or delay in the 2D tracking, whereas the power and delay are jointly included in the 3D tracking, which causes more clusters grouped as a trajectory, thus lengthening the trajectory.

\subsubsection{Occurrence Probability of Clusters}
The occurrence probability of clusters is another measure used to indicate the dynamic characteristics of clusters. In this paper, the cluster number generally ranges from 4 to 10, which are physically reasonable and easily tractable in clustering. However, we herein set the cluster number to 2-10 for precisely reaping the occurrence probability. In other words, if we use the minimum number of clusters as 4, it is no doubt that the occurrence probabilities of the first four clusters are 1. Thereupon, we statistically obtain the occurrence probabilities of all clusters, whose histogram is shown in Fig.~\ref{OP}.

Interestingly, Fig.~\ref{OP} shows that the occurrence probabilities of the first four clusters are 1, which indicates that the minimum needed number of clusters is 4, thus confirming our previous inference. For the $k$-th clusters ($k\ge5$), we found that the probability decreases with the increasing index of the cluster. Therefore, we empirically obtain the relationship between the occurrence probability and the index of cluster based on the Least-Square (LS) fitting, which is given by
\begin{equation}
	P_{\rm oc}=\left\{
	\begin{array}{ll}
		1, &  { k \le 4 }\\
		-0.115k+1.361, &  { k > 4}\\
	\end{array} \right.
\end{equation}
where the linear formulation also indicates that we have $k=11.83$ for $P_{\rm oc}=0$, which shows that the number of clusters for our measurement data should not exceed 11 to achieve a better clustering performance.

 \section{Model Implementation, Validation and Comparison}
To systematically utilize the cluster-based channel model, we will first introduce the implementation details. Subsequently, we will perform necessary validations by comparing the essential channel parameters that are calculated by the measurement and simulation channel data, respectively. Finally, we will generate cluster delay and power, constituting a general CDL model for time-varying AG channels. We will compare the AG CDL model with 3GPP Urban Macro (UMa) CDL model, aiming to provide an AG channel model to fill the gap of the current 3GPP channel model. Moreover, the result further reveals the distinct characteristics of AG channels.

\subsection{Implementation}
In this paper, we first illustrated the statistical characteristics by comprehensively demonstrating the intra-cluster and inter-cluster parameters. Then, we analyzed the corresponding cluster modeling, including the number, length, occurrence probability, and tracking, which facilitates the detailed implementation of the cluster-based channel model. Specifically, we can realize the model by the following steps. 

\subsubsection{Step 1: Set environmental parameters}
Firstly, we need to clarify the type of environment, as well as the heights, speeds, and distances of the transceiver. Note that our measurement is based on a built-up environment, therefore the model is can roughly apply to urban or suburban scenarios. Moreover, the drone height varies from ground to 30~meters, which corresponds to a low-altitude scene. 
\subsubsection{Step 2: Generate cluster delay and sub-path delay}
The number of clusters is then generated according to the Normal distribution given in Table~I. With the obtained number, the cluster delay can be obtained by Eq.~(14). Afterwards, the delay of sub-path in a cluster can be determined by adding cluster delay and delay offset generated by the Laplace distribution.
\subsubsection{Step 3: Generate cluster power and sub-path power}
With obtained cluster delay, the cluster power can be obtained by Eq.~(15). Then, the linear formula between delay and power of intra-cluster can be confirmed. Then, the power of the sub-path in a cluster can be obtained by Eq.~(9). Notably, it is necessary to update the cluster and its sub-path according to the survival length considering the time-varying channels.
\subsubsection{Step 4: Incorporate line-of-sight path}
Since the clustering process in the paper excludes the LOS path for the consideration of generality, the path can be incorporated in the rebuilt channels according to the real channel state. 
\subsubsection{Step 5: Generate channel impulse response}
Finally, the CIR can be generated by the superposition of all the generated clusters and corresponding sub-paths.

\begin{figure}[!t]
	\centering
	\subfigure[]{\includegraphics[width=3in]{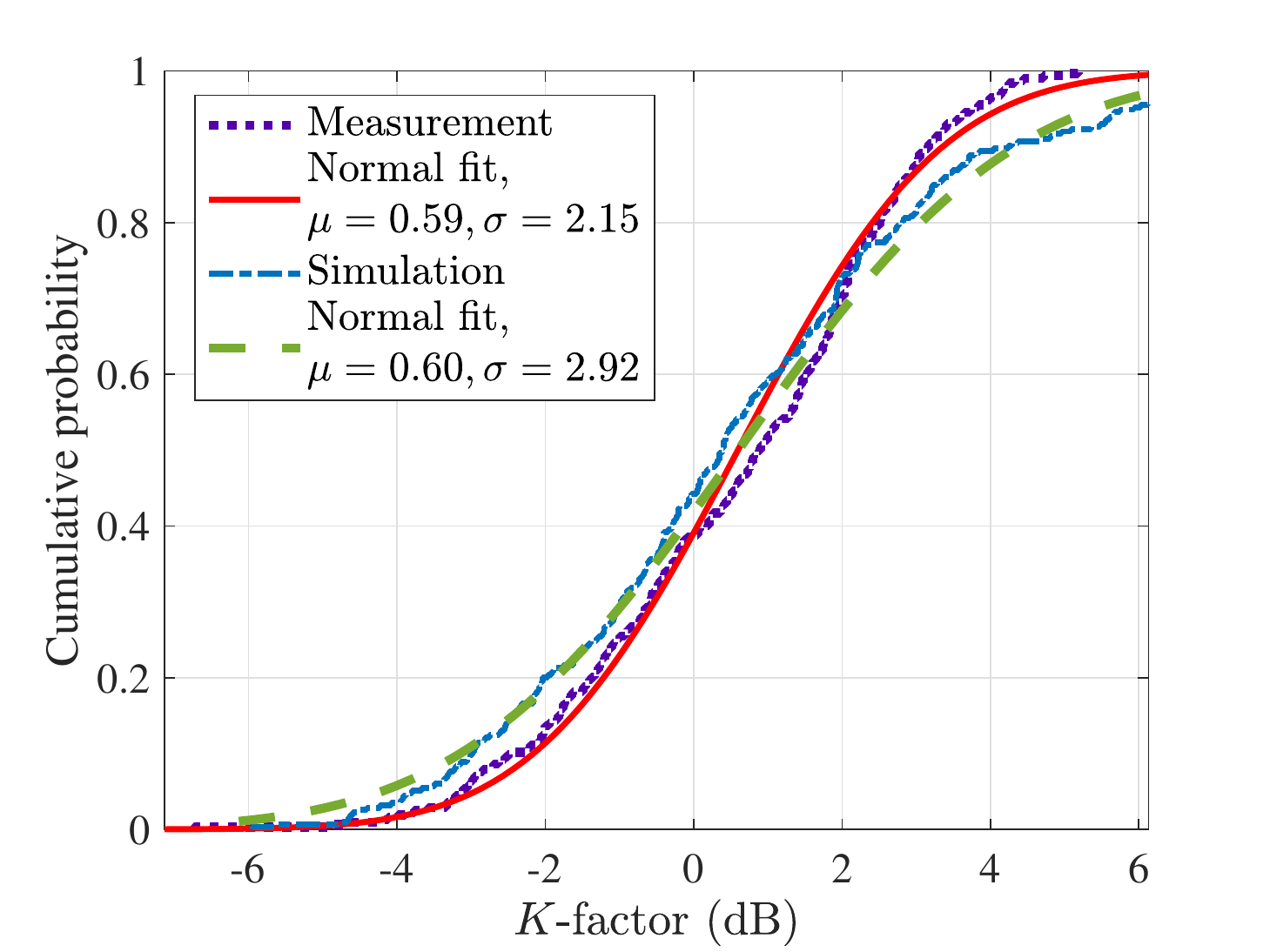}}  
	\subfigure[]{\includegraphics[width=3in]{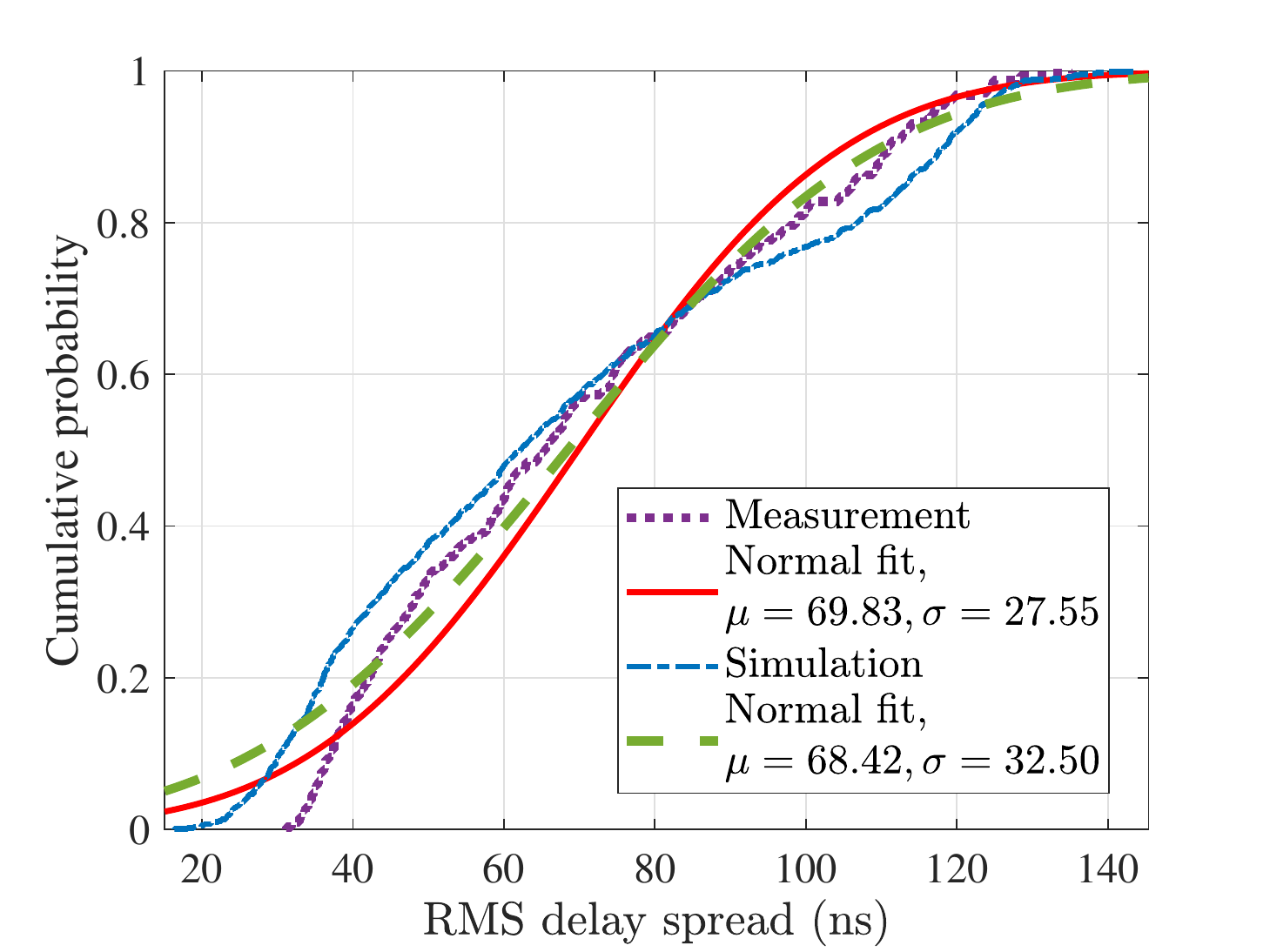}}
	\caption{Model validation: (a) Rician $K$-factor, (b) RMS delay spread.}
	\label{mvali}
\end{figure}
\subsection{Validation}
An efficient and intuitive way to verify the proposed channel model is to compare the essential parameters calculated by rebuilt CIR with measurement results, as used in prior work \cite{my1,jzl,QW}. Since Rician $K$-factor and RMS delay spread are critical channel parameters, we show the comparison results in Fig.~\ref{mvali}. Moreover, we also use popular distributions to fit the results for the mathematical interpretation. We found that simulation results present considerably good agreements with measurements. In particular, the mean values of simulations are very close to the measured results. More specifically, the mean values of Rician $K$-factor for simulation and measurement are 0.60 and 0.59 dB, respectively. For RMS delay spread, the mean values are 68.42 and 69.83 ns for the simulation and measurement, respectively. Overall, the trivial differences show the accuracy of the proposed model.

\subsection{Comparison}
The CDL model can lay out a general representation of channel profiles, thus we provide the CDL model of the AG channel, which can be constructed by generating a series of clusters with delay and power. Moreover, the UMa scenario in the 3GPP is similar to our measured environment, we incorporate its CDL model as a comparison. In addition to the LOS path, we generate a 10-cluster CDL model, where we list the absolute/scaled delay and corresponding relative power in Table II \cite{3gpp}. 

We found that the results of the AG model and 3GPP model show a certain similarity. However, there is a \emph{stronger} cluster (cluster 1) with greater power (-8.9 dB) in the AG model, which can be led by many strong MPCs existing near the LOS path, meantime they are grouped as a cluster. Moreover, there are more long-delay clusters such as clusters 8-10 in the AG model, due to the distant scatterers involving in the AG channels under the better propagation condition. These differences disclose some unique characteristics of AG propagation channels.

\begin{table}[t]
	\centering
	\caption{CDL Model Parameters} \label{table2}
	\begin{tabular}{c|c|c|c|c|c}
		\hline
		Model &  \multicolumn{3}{c}{UAV Suburban}  &  \multicolumn{2}{|c}{3GPP LOS}  \\
		\hline
		No. & \tabincell{c}{Delay \\in [ns]}&\tabincell{c}{Scaled\\delay}& \tabincell{c}{Power\\in [dB]} &\tabincell{c}{Scaled\\delay} &  \tabincell{c}{Power\\in [dB]}\\
		\hline
		LOS&0.000& 0.000  &-0.0&0.000&-0.03\\
		\hline
		1&25.67& 0.472& -8.9& 0.513& -15.8\\
		2 &33.71&0.545 & -14.9&0.544 & -18.1\\
		3&42.64& 1.079  & -18.5 &0.563& -19.8\\
		4 &50.12& 1.239 & -20.5& 0.544& -22.9 \\
		5 &60.35& 1.395 & -21.6& 0.711& -22.4\\
		6 &74.29& 1.972 & -22.5 & 1.909& -18.6\\
		7 &90.00& 2.958 & -23.3 & 1.929& -20.8\\
		8 &131.37& 3.323 & -23.8 & 1.959& -22.6\\
		9 &201.48& 3.647 & -24.3 & 2.643& -22.3\\
		10 &389.58&3.817  & -24.8 & 3.714& -25.6\\
		\hline
	\end{tabular}
\end{table}
\section{Discussion}
Although we can observe that the LOS path always exists in our measurement, while in practice, the LOS path may be absent due to blockage. We exclude the LOS since we can incorporate the LOS in a probabilistic way based on the channel state, which increases the extensibility of the model. Moreover, the angular information is unavailable due to the limitations of measurements. Nonetheless, we can easily integrate the angles in the model based on the complete cluster-based modeling process. For example, we can conduct multi-dimensional clustering by slightly altering the expressions of MCD in the KPM algorithm and the Euclidean distance in the KM method. Therefore, the proposed methodology in the paper is essential for both the UAV-based AG channel and future model extension with multi-dimensional MPC information.

\section{Conclusion}
In this paper, we conducted the cluster-based characterization and modeling for UAV-based time-varying A2G channels. We first provided the statistical characteristics by comprehensively analyzing the intra-cluster and inter-cluster parameters. We found that at least 4 clusters should be used for accurately describing the AG clustered channel. More invaluably, we developed several novel methodologies for cluster characterization and modeling, where the rectangle method and double exponential function are useful to analyze the intra-cluster and inter-cluster power decay characteristics, respectively. It is found that the average power decay degree is 0.53 dB/ns for intra-clusters. Moreover, the clustering-based tracking was proposed for the first time to quantify the dynamic feature of clusters, where we found the average survival length is 5.46~m in the 3D tracking result. Then, the cluster-based channel model is validated and compared to show its accuracy and generality. Finally, we discussed insightful considerations and limitations of the work, which paves the way for future work.


\begin{thebibliography}{99}
\bibitem{r1}
Y. Zeng, Q. Wu and R. Zhang, ``Accessing from the sky: a tutorial on UAV communications for 5G and beyond,'' \emph{Proceedings of the IEEE}, vol. 107, no. 12, pp. 2327-2375, Dec. 2019.

\bibitem{huawei}
Huawei, ``Huawei tests world's first 5G base station on drones,'' Jan. 2020. [Online]. Available: https://cntechpost.com/2020/01/02/huawei-tests-worlds-first-5g-base-station-on-drones/

\bibitem{nokia}
Nokia, ``F-Cell technology from Nokia Bell Labs revolutionizes small cell deployment by cutting wires, costs and time,'' Oct. 2016. [Online]. Available: https://www.nokia.com/about-us/news/releases/2016/10/03/f-cell-technology-from-nokia-bell-labs-revolutionizes-small-cell-deployment-by-cutting-wires-costs-and-time/

\bibitem{att1}
AT\&T, ``When COWs fly: AT\&T sending LTE signals from drones,'' Feb. 2017. [Online]. Available: https://about.att.com/innovationblog/cows\_fly/

\bibitem{zc1}
Z. Cui, C. Briso-Rodr\'iguez, K. Guan, Z. Zhong and F. Quitin, ``Multi-frequency air-to-ground channel measurements and analysis for UAV communication systems,'' \emph{IEEE Access}, vol. 8, pp. 110565-110574, Jun. 2020.

\bibitem{cb1}
Z. Cui, C. Briso-Rodr\'iguez, K. Guan, C. Calvo-Ram\'irez, B. Ai and Z. Zhong, ``Measurement-based modeling and analysis of UAV air-ground channels at 1 and 4 GHz,'' \emph{IEEE Antennas Wireless Propag. Lett.}, vol. 18, no. 9, pp. 1804-1808, Sept. 2019.

\bibitem{dw1}
D. W. Matolak and R. Sun, ``Air-ground channel characterization for unmanned aircraft systems-part i: Methods, measurements, and models for over-water settings,'' \emph{IEEE Trans. Veh. Technol.}, vol. 66, no. 1, pp. 26-44, Jan. 2017.

\bibitem{dw2}
R. Sun and D. W. Matolak, ``Air-ground channel characterization for unmanned aircraft systems-part ii: Hilly and mountainous settings,'' \emph{IEEE Trans. Veh. Technol.}, vol. 66, no. 3, pp. 1913-1925, Mar. 2017.

\bibitem{dw3}
D. W. Matolak and R. Sun, ``Air-ground channel characterization for unmanned aircraft systems-part iii: The suburban and near-urban environments,'' \emph{IEEE Trans. Veh. Technol.}, vol. 66, no. 8, pp. 6607-6618, Aug. 2017.

\bibitem{qz}
Q. Zhu, K. Jiang, X. Chen, W. Zhong and Y. Yang, ``A novel 3D non-stationary UAV-MIMO channel model and its statistical properties,'' \emph{China Commun.}, vol. 15, no. 12, pp. 147-158, Dec. 2018.

\bibitem{tsr1}
T. S. Rappaport, \emph{Wireless communications: principles and practice}, Upper Saddle River, N.J.: Prentice Hall PTR, 2002.

\bibitem{rh1}
R. He \emph{et al.}, ``A kernel-power-density-based algorithm for channel multipath components clustering,'' \emph{IEEE Trans. Wirel. Commun.}, vol. 16, no. 11, pp. 7138-7151, Nov. 2017.

\bibitem{tw1}
T. Wu, X. Yin and J. Lee, ``A novel power spectrum-based sequential tracker for time-variant radio propagation channel,'' \emph{IEEE Access}, vol. 8, pp. 151267-151278, Aug. 2020.

\bibitem{hs1}
H. Suzuki, ``A statistical model for urban radio propagation,'' \emph{IEEE Trans. Commun.}, vol. COM-25, no. 7, pp. 673-680, Jul. 1977.

\bibitem{sv}
A. A. M. Saleh and R. A. Valenzuela, ``A statistical model for indoor multipath propagation,'' \emph{IEEE J. Sel. Areas Commun.}, vol. SAC-5, no. 2, pp. 128–137, Feb. 1987.

\bibitem{cost}
L. Liu \emph{et al}., ``The COST 2100 MIMO channel model,'' \emph{IEEE Wireless Commun.}, vol. 19, no. 6, pp. 92–99, Dec. 2012.

\bibitem{3gpp}
``Study on channel model for frequencies from 0.5 to 100 GHz,
V15.0.0,'' 3GPP, Sophia Antipolis, France, Rep. TR 38.901,
Jun. 2018. [Online]. Available: https://portal.3gpp.org/desktopmodules/
Specifications/SpecificationDetails.aspx?specificationId=3173

\bibitem{winner}
 J. Meinilä, P. Kyösti, T. Jämsä, and L. Hentilä, ``WINNER II channel models,''  in \emph{Radio Technologies and Concepts for IMT-Advanced}. Hoboken, NJ, USA: Wiley, 2009, pp. 39–92.

 \bibitem{my1}
M. Yang \emph{et al.}, ``A cluster-based three-dimensional channel model for vehicle-to-vehicle communications,'' \emph{IEEE Trans. Veh. Technol.}, vol. 68, no. 6, pp. 5208-5220, Jun. 2019.

 \bibitem{ch}
C. Huang, A. F. Molisch, R. He, R. Wang, P. Tang and Z. Zhong, ``Machine-learning-based data processing techniques for vehicle-to-vehicle channel modeling,'' \emph{IEEE Commun. Mag.}, vol. 57, no. 11, pp. 109-115, Nov. 2019.

\newpage


\vspace{10in}
\bibitem{hj}
H. Jiang, W. Ying, J. Zhou and G. Shao, ``A 3D wideband two-cluster channel model for massive MIMO vehicle-to-vehicle communications in semi-ellipsoid environments,'' \emph{IEEE Access}, vol. 8, pp. 23594-23600, Jan. 2020.

\bibitem{ccr}
C. Calvo-Ram\'irez, Z. Cui, C. Briso, K. Guan and D. W. Matolak, ``UAV air-ground channel ray tracing simulation validation,'' in \emph{Proc. IEEE/CIC International Conference on Communications in China (ICCC Workshops)}, Beijing, China, Jun. 2018, pp. 122-125.


\bibitem{jl1}
J. Lee, ``Cluster-based millimeter-wave outdoor-to-indoor propagation characteristics based on 32 GHz measurement analysis,'' \emph{IEEE Antennas Wireless Propag. Lett.}, vol. 20, no. 1, pp. 73-77, Jan. 2021.

\bibitem{um}
U. Maulik and S. Bandyopadhyay,  ``Performance evaluation of some clustering algorithms and validity indices,'' \emph{IEEE Trans. Pattern Anal. Mach. Intell.}, vol. 24, no. 12, pp. 1650-1654, Dec. 2002.

\bibitem{zyh}
Z. Huang, J. Rodríguez-Piñeiro, T. Domínguez-Bolaño, X. Cai and X. Yin, "Empirical dynamic modeling for low-altitude UAV propagation channels,'' \emph{ IEEE Trans. Wireless Commun.}, Early Access. doi: 10.1109/TWC.2021.3065959.


\bibitem{ch1}
C. Huang, A. F. Molisch, Y. Geng, R. He, B. Ai and Z. Zhong, ``Trajectory-joint clustering algorithm for time-varying channel modeling,'' \emph{IEEE Trans. Veh. Technol.}, vol. 69, no. 1, pp. 1041-1045, Jan. 2020.

\bibitem{dwm}
 DWM1001 DataSheet, version 2.08, Decawave, Ltd, 2016.

 \bibitem{zc2}
Z. Cui, C. Briso-Rodr\'iguez, K. Guan, \.I. G\"uven\c c and Z. Zhong, ``Wideband air-to-ground channel characterization for multiple propagation environments,'' \emph{IEEE Antennas Wireless Propag. Lett.}, vol. 19, no. 9, pp. 1634-1638, Sept. 2020.

 \bibitem{xc1}
X. Cai \emph{et al.}, ``An empirical air-to-ground channel model based on passive measurements in LTE,'' \emph{IEEE Trans. Veh. Technol.}, vol. 68, no. 2, pp. 1140-1154, Feb. 2019.

\bibitem{xw}
X. Wu et al.,``60-GHz millimeter-wave channel measurements and modeling for indoor office environments,'' \emph{IEEE Trans. Antennas Propag.}, vol. 65, no. 4, pp. 1912–1924, Apr. 2017.

 \bibitem{sage}
B. H. Fleury, M. Tschudin, R. Heddergott, D. Dahlhaus, and K. I. Pedersen, ``Channel parameter estimation in mobile radio environments using the SAGE algorithm,'' \emph{IEEE J. Sel. Areas Commun.}, vol. 17, no. 3, pp. 434–450, Mar. 1999.

 \bibitem{kg1}
M. Kim, S. Kishimoto, S. Yamakawa and K. Guan, ``Millimeter-wave intra-cluster channel model for in-room access scenarios,'' \emph{IEEE Access}, vol. 8, pp. 82042-82053, Apr. 2020.

\bibitem{cir}
 W. Khawaja, O. Ozdemir, F. Erden, I. Guvenc and D. W. Matolak, ``UWB air-to-ground propagation channel measurements and mdeling using UAVs,'' in \emph{Proc. IEEE Aerospace Conference}, Big Sky, MT, USA, 2019, pp. 1-10.

\bibitem{kpm}
N. Czink, P. Cera, J. Salo, E. Bonek, J.-P. Nuutinen, and J. Ylitalo, ``A framework for automatic clustering of parametric MIMO channel data including path powers,'' in \emph{Proc. IEEE VTC}, Sep. 2006, pp. 1-5.

\bibitem{km}
J. MacQueen, ``Some methods for classification and analysis of multivariate observations,'' in \emph{Proc. BSMSP}, 1967, pp. 281–297.

\bibitem{si}
P. J. Rousseeuw, ``Silhouettes: A graphical aid to the interpretation and validation of cluster analysis,'' \emph{J. Comput. Appl. Math.}, vol. 20, no. 1, pp. 53–65, 1987.

\bibitem{db}
D. L. Davies and D. W. Bouldin, ``A cluster separation measure,'' \emph{ IEEE Trans. Pattern Anal. Mach. Intell.}, vol. PAMI-1, no. 2, pp. 224-227, Apr. 1979.

\bibitem{jzl}
J. Li, B. Ai, R. He, M. Yang, Z. Zhong and Y. Hao, ``A cluster-based channel model for massive MIMO communications in indoor hotspot scenarios,'' \emph{ IEEE Trans. Wireless Commun.}, vol. 18, no. 8, pp. 3856-3870, Aug. 2019.

\bibitem{ch2}
C. Huang, R. He, Z. Zhong, Y. Geng, Q. Li and Z. Zhong, ``A novel tracking-based multipath component clustering algorithm,'' \emph{IEEE Antennas Wirel. Propag. Lett.}, vol. 16, pp. 2679-2683, Aug. 2017.


\bibitem{QW}
Q. Wang \emph{et al.}, ``Time-variant cluster-based channel modeling for V2V communications,'' in \emph{Proc. IEEE ICC}, 2018, pp. 1-6.




\end{thebibliography}
\end{document}